\documentclass[preprint,preprintnumbers,prd,amsmath,amssymb,nofootinbib,tightenlines]{revtex4}
\usepackage{graphicx}% Include figure files
\usepackage{dcolumn}% Align table columns on decimal point
\usepackage{bm}% bold math
\usepackage{slashed}
\def\op{{\bf P}}

\def\ot{{\bf T}}
\def\cp{{\bf CP}}
\def\cpt{{\bf CPT}}

\begin{document}

\title{\boldmath%
${\mathbf T}$-odd Momentum Correlation in Radiative ${\mathbf \beta}$ Decay 
\unboldmath}

\author{Susan Gardner and Daheng He}

\affiliation{Department of Physics and Astronomy, University of Kentucky, 
Lexington, KY 40506-0055
}

%\date{\today}

\begin{abstract}
We consider neutron radiative 
$\beta$ decay, $n\rightarrow pe^-\bar{\nu}_e\gamma$, 
and compute the \ot-odd momentum correlation in the decay rate 
characterized by the kinematical variable 
$\xi = {\mathbf l}_\nu\cdot({\mathbf l}_e\times{\mathbf k})$ 
arising from electromagnetic final-state interactions in the Standard Model. 
Our expression for the corresponding \ot-odd asymmetry $A_\xi^{\rm SM}$ 
is exact in ${\cal O}(\alpha)$ up to terms of recoil order, and we evaluate 
it numerically under various kinematic conditions. 
Noting the universality of the V-A law in the absence of recoil-order terms, we 
retain the parametric dependence on masses and coupling constants
throughout, so 
that our results serve as a template for the computation of
$A_\xi^{\rm SM}$ in 
allowed nuclear radiative $\beta$ decays 
and hyperon radiative $\beta$ decays as well. 
%We consider the pattern of $A_\xi^{\rm SM}$ which appears 
%in nuclear decays and show that $A_\xi^{\rm SM}$ can be suppressed in 
%particular cases, facilitating searches for new sources of \cp-violation in such processes. 

\end{abstract}

\maketitle

\section{Introduction}

Radiative $\beta$ decay offers the opportunity of studying \ot-odd momentum correlations
which do not appear in ordinary $\beta$ decay~\cite{Jackson:1957zz}. We consider a 
correlation characterized by the kinematical 
variable $\xi = {\mathbf l}_\nu\cdot({\mathbf l}_e\times{\mathbf k})$, 
% old def
%$\xi = {\mathbf l}_e\cdot({\mathbf l}_v\times{\mathbf k})$, 
so that it is both parity \op \ and naively time-reversal \ot \ odd 
but independent of the particle spin. Its spin independence renders it distinct from searches for 
permanent electric-dipole moments (EDMs) of neutrons and nuclei. 
The inability of the Standard Model (SM) to explain the cosmic baryon asymmetry
prompts the search for sources of \cp \ violation which do not appear within it
and which are not constrained by other experiments. 
A triple momentum correlation in radiative $\beta$ decay is one such example, as we shall illustrate; 
under the \cpt \ theorem, \ot \ violation is linked to \cp \ violation. 
A decay correlation, however, can be, 
by its very nature, only ``naively'' or ``pseudo'' \ot \ odd, that is, 
only motion-reversal odd. 
 As a result, 
although the appearance of a \ot-odd decay correlation can be engendered by 
sources of \cp \ violation beyond the Standard Model, it can also be generated
without fundamental \ot \ or \cp \ violation. 
In this paper we compute the size of the \ot-odd momentum correlation 
in radiative $\beta$ decay simulated by 
electromagnetic final-state interactions in the SM~\cite{sachs}. 
This is crucial to establishing a baseline in the search for new sources of 
\cp \ violation 
in such processes. Our work is motivated 
in large part by the determination that pseudo-Chern-Simons 
terms appear in SU(2)${}_{\rm L}\times$U(1) gauge theories at 
low energies -- and that they can 
impact low-energy weak radiative processes 
involving baryons~\cite{Harvey:2007rd,Harvey:2007ca,Hill:2009ek}. 
In the SM such pseudo-Chern-Simons interactions are \cp \ conserving, 
but considered broadly they are not, so that searching 
for the \op- and \ot-odd effects that \cp-violating 
interactions of pseudo-Chern-Simons form would engender
offers a new window on physics beyond the SM~\cite{svgdh1}. 

Searches for \ot-violating decay correlations in neutron and nuclear $\beta$ decay
have a long history. The best experimental limits are 
on the so-called $D$ term, which appears as the 
triple correlation $D {\mathbf S}\cdot ({\mathbf l}_e \times {\mathbf l}_\nu)$, 
where ${\mathbf S}$ is the polarization of 
the decaying particle~\cite{Mumm:2011nd,calaprice}.
These limits still greatly exceed 
the size of the $D$ correlation 
expected from SM final-state interactions~\cite{Callan:1967zz,Ando:2009jk}.  
Radiative $\beta$ decay offers the possibility of forming a \ot-odd correlation
from momenta alone; to our knowledge such a possibility was first considered in the context of 
$K_{l3\gamma}^+$ decay~\cite{braguta}. The \ot-odd asymmetry computed in  
Ref.~\cite{braguta} from electromagnetic final-state interactions 
has recently been recalculated and is in significant disagreement 
with the earlier result~\cite{khriprud}. 
%by Khriplovich and Rudenko

In this paper we evaluate the \ot-odd asymmetry in radiative $\beta$ decay from 
electromagnetic radiative corrections in the SM 
and focus on the neutron case: 
$n(p_n)\rightarrow p(p_p) +  e^-(l_e) + \bar{\nu}_e(l_\nu) + \gamma(k)$.  
The motion-reversal-odd terms in the decay rate, 
which mimic the appearance of \ot \ violation, are
engendered by the interference of the tree-level amplitude with the imaginary part of the 
${\cal O}(\alpha)$ corrected amplitude, 
which is determined by the physical two-particle cuts and
hence mediated by the scattering of particles
on their mass shells~\cite{Cutkosky:1960sp,Callan:1967zz,Okun:1967ww}. 
In what follows, we detail the computation of 
the interference terms and their components, as well as the 
resulting numerical integration over the allowed phase space to yield 
the \ot-odd asymmetry $A_\xi^{\rm SM}$. 
Our results are exact in ${\cal O}(\alpha)$ up to corrections of 
recoil order, namely, up to terms of ${\cal O}(\varepsilon/M)$, where $\varepsilon$ 
is an energy scale which is small with respect to the nucleon mass $M$. 
This certitude is guaranteed by 
%the small energy release of the 
the small $Q$ value of the decay, so that $\varepsilon \ll M$, and by 
Low's theorem~\cite{Low:1958sn}. 
The natural scale of hadron excitations is set by 
the pion mass $m_\pi$; consequently, in neutron radiative
$\beta$ decay  $\varepsilon \ll m_\pi$ as well, and nonelectromagnetic
final-state interactions cannot contribute to the physical two-particle cuts. 
This is in contradistinction to $K_{l3\gamma}^+$ decay 
for which such contributions are appreciable, 
albeit relatively small~\cite{Muller:2006gu}. 
We relegate intermediate results essential for our final results but yet 
nonessential to the flow of our discussion to Appendixes. Since we neglect all 
terms of recoil order, our results are relevant to the computation
of $A_\xi^{\rm SM}$ in nuclear and hyperon radiative $\beta$ decays as well. 
%We consider, in particular, the pattern in $A_\xi^{\rm SM}$ predicated by 
%variations in the $Q$-value of 
%%energy released in 
%the decay as well as by the nuclear dependence of the
%axial-vector coupling constant.
We assess the size of undetermined corrections
%, which are likely larger
%in nuclear decays, 
before offering a final summary of our results. 

\section{Formalism}
\label{formal}

\begin{figure}
\begin{center}
\includegraphics[scale=0.45]{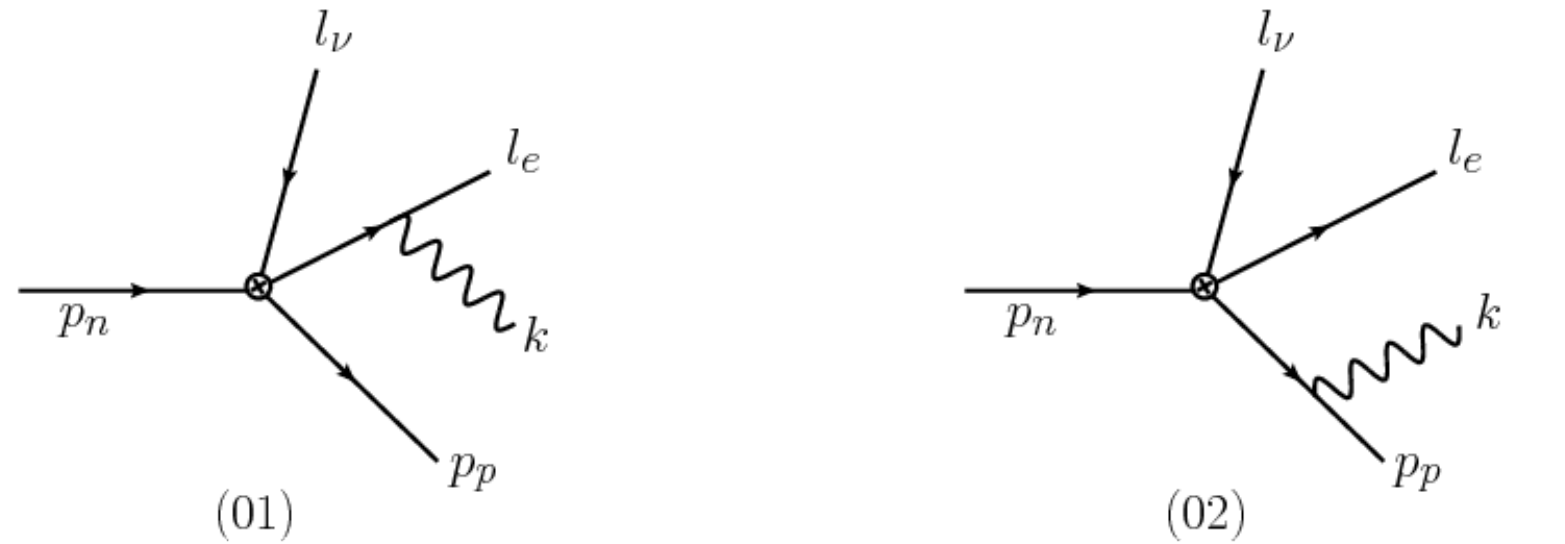}
\caption{Contributions to $n(p_n)\to p(p_p) +  e^-(l_e) + {\bar\nu}_e(l_\nu) + \gamma(k)$ 
up to corrections of recoil order. 
The effective weak vertex is denoted by $\otimes$ and is controlled by the
Fermi constant $G_{\rm F}$. 
The diagram enumeration is utilized in our calculation of the \ot-odd asymmetry. 
}
\label{fig:tree}
\end{center}
\end{figure}

We work in a simultaneous expansion in the electromagnetic coupling constant $e$
and in $\varepsilon/M$, 
so that the leading contributions to neutron radiative $\beta$ decay 
are from the diagrams in Fig.~\ref{fig:tree}. 
At this order the baryons are effectively structureless, and 
the contributions arise from bremsstrahlung off the charged particle legs of ordinary $\beta$ decay, 
yielding a gauge-invariant result~\cite{Low:1958sn}. 
Employing the notation and conventions of Ref.~\cite{PS}, the decay amplitude is:
\begin{equation}
{\cal M}_{\rm tree}={\cal M}_{01}+{\cal M}_{02}\,,
\end{equation}
with~\cite{gaponov,bgmz}
% drop i in bgmz
\begin{eqnarray}
{\cal M}_{01}(l_e,k,p_p)&=&\frac{eg_VG_F}{\sqrt{2}}\bar{u}_e(l_e)\frac{2l_e\cdot\epsilon^\ast+\slashed{\epsilon}^\ast\slashed{k}}{2l_e\cdot k}\gamma_\rho(1-\gamma_5)v_\nu(l_\nu)\bar{u}_p(p_p)\gamma^\rho(1-\lambda\gamma_5)u_n(p_n) \,, 
\label{calM0} \\
{\cal M}_{02}(l_e,k,p_p)&=&-\frac{eg_VG_F}{\sqrt{2}}\bar{u}_e(l_e)\gamma_\rho(1-\gamma_5)v_\nu(l_\nu)\bar{u}_p(p_p)\frac{2p_p\cdot\epsilon^\ast+\slashed{\epsilon}^\ast\slashed{k}}{2p_p\cdot k}\gamma^\rho(1-\lambda\gamma_5)u_n(p_n) \,,
\end{eqnarray}
where 
$\epsilon^\mu$ is the photon polarization vector and 
$\lambda \equiv g_A/g_V$, noting that $g_V$ and $g_A$ are the vector and axial-vector
weak coupling constants of the nucleon, respectively. 
We explicitly include the arguments 
in the momenta $l_e$, $k$, and $p_p$ for later convenience. 

The branching ratio and photon energy spectrum for this process
have been computed previously~\cite{gaponov,bgmz}. The expressions which follow from 
Eq.~(\ref{calM0}) are consistent 
%compare favorably 
with the experimental results~\cite{nico,cooper}. 
The next-to-leading order terms in the small-scale expansion, i.e., 
those of ${\cal O}(\varepsilon/M)$, have been computed in heavy-baryon chiral perturbation theory 
and are no larger than ${\cal O}(0.5\%)$ of
the leading-order result~\cite{bgmz} -- this is some 20 times smaller 
than the current experimental sensitivity~\cite{cooper}. In what follows we neglect
all recoil-order terms and consider the ${\cal O}(\alpha)$ corrections
to the amplitude of Eq.~(\ref{calM0}). For future reference, 
employing lepton and hadron tensors, we note that~\cite{bgmz}
\begin{equation}
\sum_{\rm spins}|{\cal M}_0|^2=\frac{e^2g^2_VG^2_F}{2}
\left(\frac{1}{(l_e\cdot k)^2}L^{\rm ee}_{\rho\delta}H^{\rho\delta}+\frac{1}{M_p^2\omega^2}L^{\rho\delta}H^{\rm ee}_{\rho\delta}-\frac{1}{M_p\omega(l_e\cdot k)}M^{\rm ee,mixed}\right) \,,
\label{calM0sq}
\end{equation}
where $M_n$, $M_p$, and $\omega$ refer to the neutron mass, the proton mass, and the
photon energy, respectively, and 
\begin{eqnarray}
L^{\rm ee}_{\rho\delta}H^{\rho\delta}&=&-64M_nM_p\left(m^2_e-l_e\cdot k\right)\left((1+3\lambda^2)
E_\nu(E_e+\omega)+(1-\lambda^2)({\bf l}_e\cdot{\bf l}_\nu+{\bf l}_\nu\cdot{\bf k} )\right) 
\,,
\nonumber \\ 
L^{\rho\delta}H^{\rm ee}_{\rho\delta}&=&-64M_nM_p^3\left((1+3\lambda^2)E_\nu E_e+(1-\lambda^2)
{\bf l}_e\cdot{\bf l}_\nu \right) \,,
\label{tensors} \\
M^{\rm ee,mixed}&=&-64M_nM_p^2
\left((1+3\lambda^2)E_\nu(2E_e^2+E_e\omega-l_e\cdot k)+(1-\lambda^2)
E_e (2 {\bf l}_e\cdot{\bf l}_\nu+ {\bf l}_\nu\cdot {\bf k})\right) \,, \nonumber
\end{eqnarray}
with $m_e$ the electron mass. 
In realizing the amplitudes from the Feynman rules 
we impose $\epsilon_i (k)\cdot k=0$ 
for each polarization state $i$ of a real photon with momentum $k$, 
noting $\sum_{i=1,2}\epsilon_i^\ast (k)\cdot \epsilon_i (k) = -2$. 
To effect the subsequent photon polarization sums, however, we 
employ QED gauge invariance and make the replacement 
$\sum_{i=1,2} \epsilon_i^\mu(k) \epsilon_i^{\nu\,\ast}(k) \longrightarrow - g^{\mu \nu}$
throughout, without any supplemental conditions. 

Denoting the ${\cal O}(\alpha)$ correction to the amplitude by ${\cal M}_{\rm loop}$ the 
amended decay rate is determined by 
\begin{equation}
|{\cal M}|^2=
|{\cal M}_{\rm tree}|^2 + 
{\cal M}_{\rm tree}\cdot{{\cal M}^\ast_{\rm loop}} + {{\cal M}_{\rm loop}}\cdot{{\cal M}^\ast_{\rm tree}}+ {\cal O}(\alpha^2) 
\,.
\end{equation}
The \ot-odd triple momenta correlation 
$\xi = {\mathbf l}_\nu\cdot({\mathbf l}_e\times{\mathbf k})$ in the decay rate 
can arise from the interference between the tree-level 
amplitude ${\cal M}_{\rm tree}$ and the anti-Hermitian parts 
of the one-loop corrections to it, so that ultimately the interference term 
$\sum_{\rm spins} (2 {\rm Re}({\cal M}_{\rm tree} {\cal M}^\ast_{\rm loop}))$ 
contains terms linear in $\xi$. 
Since we consider the decay and detection of unpolarized particles exclusively, 
$\sum_{\rm spins} |{\cal M}|^2_{\rm \ot \ odd}$ is 
indeed characterized by terms linear in $\xi$. 
Evidently the induced asymmetry is suppressed by a factor of 
$\alpha\equiv e^2/4\pi \sim 1/137$; explicit computation shows it to be much smaller still. 

Before turning to the computation of $|{\cal M}|^2_{\rm \ot \ odd}$ let us 
consider its relation to a measurable quantity. 
Following Ref.~\cite{braguta}, we define a \ot-odd asymmetry $A_\xi$, namely, 
\begin{equation}
A_\xi=\frac{N_+-N_-}{N_++N_-}\,,
\end{equation}
where $N_+$ is defined as 
the total number of decay events with positive $\xi$, 
and $N_-$ is defined as the number of events with negative $\xi$. 
Specifically, we compute
\begin{equation}
A_\xi=\frac{\Gamma_+-\Gamma_-}{\Gamma_++\Gamma_-} \,,
\label{asym}
\end{equation}
where $\Gamma_{\pm}$ contains an integral of
$|{\cal M}|^2$
over the region of phase space with 
$\xi\stackrel{>}{{}_<} 0$, respectively; the numerator is nonzero if and only if
$|{\cal M}|^2_{\rm \ot \ odd}$ is nonzero. 
Working to corrections of ${\cal O}(\varepsilon/M)$, the 
neutron radiative $\beta$-decay rate $\Gamma$ in the neutron rest frame is 
\begin{equation}
\Gamma=\frac{1}{8M_n}\frac{1}{(2\pi)^8}
\int |{\mathbf l}_e|d{E_e}d{\Omega_e} \omega d{\omega} d{\Omega_k}d{\Omega_\nu}
\frac{\Theta(M_n-E_e-E_\nu-\omega)E_\nu}{4M_n} 
\left(
\frac{1}{2} 
\sum_{\rm spins} |{\cal M}|^2\right)\Bigg|_{p_p,\,E_\nu}
\,,
\end{equation}
where $p_p=M_n-l_e-l_\nu-k$ and $E_\nu = M_n - M_p - E_e - \omega$
are fixed throughout. 
The precise form of $\Gamma_\pm$ depends on the concrete choice of coordinate system. 
Choosing the direction of the electron momentum ${\mathbf l}_e$ 
as the ${\mathbf z}$ direction and letting ${\mathbf k}$ and ${\mathbf l}_e$ 
fix the $\mathbf{x}$-$\mathbf{z}$ plane, then 
under this specific choice $\xi>0$ corresponds to $\phi_\nu\in [0,\pi]$ and $\xi<0$ 
corresponds to $\phi_\nu\in [\pi,2\pi]$. Thus we define 
\begin{eqnarray}
\nonumber 
\Gamma_{+}(\omega^{\rm min})\equiv\frac{1}{16M_n^2(2\pi)^6}
\int^{\omega^{\rm max}}_{\omega^{\rm min}} 
\omega d\omega\int^{E_e^{\rm max}(\omega)}_{m_e}|{\mathbf l}_e|
dE_e\int^{\rm c}_{\rm -c} dx_k\int^{1}_{-1}dx_{\nu}\int^\pi_0d\phi_\nu E_\nu  \\
\times 
\left(
\frac{1}{2} 
\sum_{\rm spins} |{\cal M}|^2\right)\Bigg|_{p_p,\,E_\nu} \,,
\end{eqnarray}
and 
\begin{eqnarray}
\nonumber \Gamma_{-}(\omega^{\rm min})\equiv\frac{1}{16M_n^2(2\pi)^6}
\int^{\omega^{\rm max}}_{\omega^{\rm min}}\omega d\omega\int^{E_e^{\rm max}(\omega)}_{m_e}
|{\mathbf l}_e|dE_e\int^{\rm c}_{\rm -c} dx_k \int^{1}_{-1}dx_{\nu}\int^{2\pi}_\pi d\phi_\nu E_\nu  \\
\times 
\left(
\frac{1}{2} 
\sum_{\rm spins} |{\cal M}|^2\right)\Bigg|_{p_p,\,E_\nu} \,,
\end{eqnarray}
where $E_e^{\rm max} = M_n - M_p - \omega$, $\omega^{\rm max} = M_n - M_p - m_e$, 
and $\omega^{\min}$ is determined by the threshold energy 
%resolution 
of the detector. 
In our computation of $|{\cal M}|^2$ we set $M_n = M_p = M$ in terms which would
yield corrections beyond leading order in the recoil expansion. 
We limit the integration over $x_k$ to the range $[-{\rm c},{\rm c}]$; 
we discuss this as well as our choice for ${\rm c}$ in Sec.~\ref{res}.

\section{Computation of $\sum_{\rm spins} \mathbf{|{\cal M}|^2_{\ot \ odd}}$ in leading order}

\begin{figure}
\begin{center}
\includegraphics[scale=0.45]{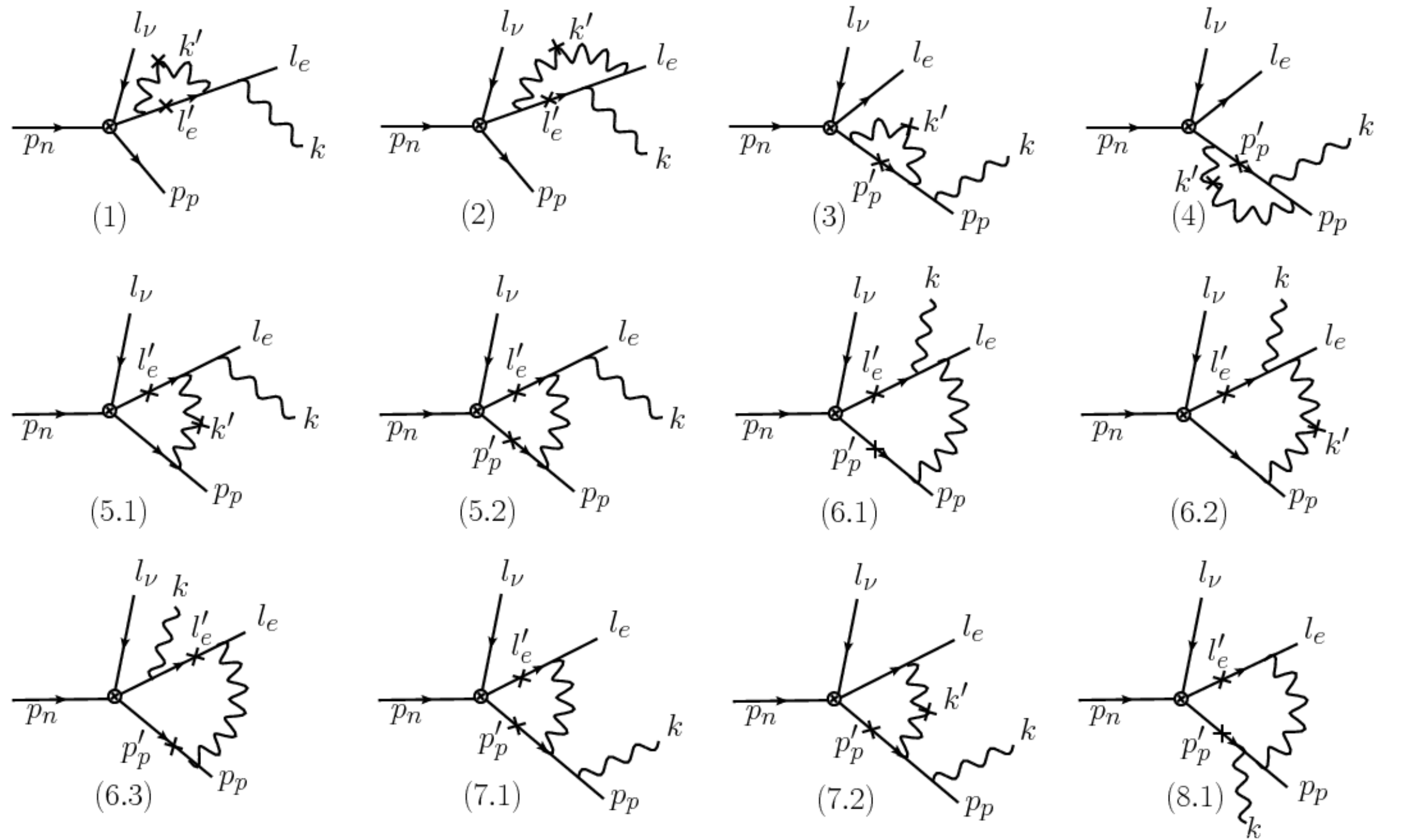}
\includegraphics[scale=0.45]{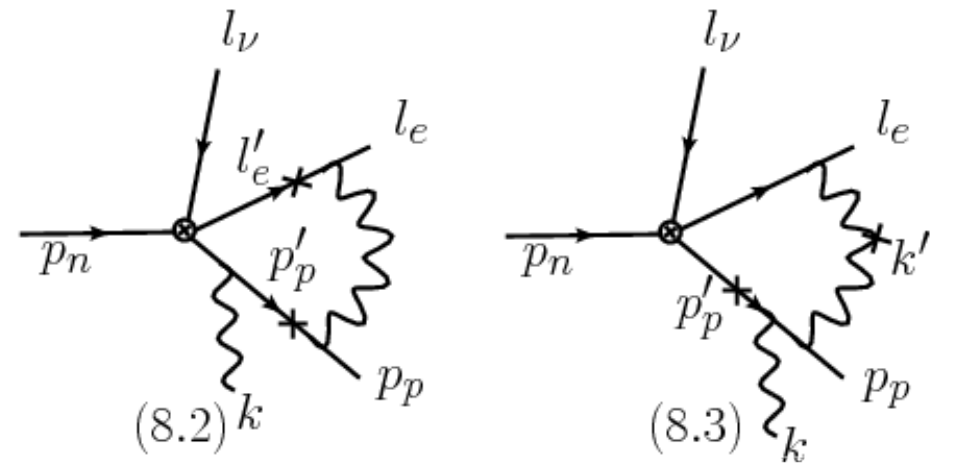}
\caption{All two-particle cut contributions 
to $n(p_n)\to p(p_p) + e^-(l_e) + {\bar\nu}_e(l_\nu) + \gamma(k)$ 
which appear in ${\cal O}(\alpha)$ up to corrections of recoil order, 
using the syntax of Fig.~\ref{fig:tree}. A ``$\times$'' means that the intermediate 
particle has been put on its mass shell; two such symbols define the Cutkosky cut. 
The diagram enumeration is utilized in our calculation of the \ot-odd asymmetry. Note
that the first number selects a particular Feynman diagram, and the second determines
the particular two-particle physical cut in that diagram.  
}
\label{fig:allcuts}
\end{center}
\end{figure}

To compute the \ot-odd pieces, 
we need to obtain the anti-Hermitian parts of the 
one-loop diagrams ${\rm Im} ({\cal M}_{\rm loop})$. We do this by performing 
 ``Cutkosky cuts''~\cite{Cutkosky:1960sp}, which means we simultaneously put intermediate 
particles in the loops on their mass shells in all physically allowed ways and then 
perform the relevant intermediate phase-space integrals and spin sums. 
Graphically speaking, after imposing the cuts, the anti-Hermitian part of a one-loop diagram can be 
viewed as the product of two physical tree-level processes. We have 
\begin{equation}
{\rm Im}({\cal M}_{\rm loop})=\frac{1}{8\pi^2}\sum_n \int d\rho_n \sum_{s_n}{\cal M}_{fn}{{\cal M}^\ast_{in}}
=\frac{1}{8\pi^2}\sum_n \int d\rho_n \sum_{s_n}{\cal M}_{fn}{{\cal M}_{ni}} \,,
\end{equation}
where $\sum_{n}$ refers to the summation over 
all the possible cuts of the one-loop diagrams and $\int d\rho_n$ and $\sum_{s_n}$ refer to the intermediate
phase space integration and spin sums, respectively, for a cut which yields state $n$. 
The matrix elements ${\cal M}_{ni}$ and 
${\cal M}_{fn}$ refer to the two tree-level diagrams after a physical cut. 
After excluding the physically unacceptable cuts, 14 cut diagrams remain, and  
they are illustrated in Fig.~\ref{fig:allcuts}. 
We evaluate them explicitly. The momenta labeled as $k',\, l_e'$, and $p_p'$ refer to momenta 
of intermediate particles. In performing the Cutkosky cuts, 
each particle in a pair of particles 
is put on its own mass shell. 

\begin{figure}
\begin{center}
\includegraphics[scale=0.4]{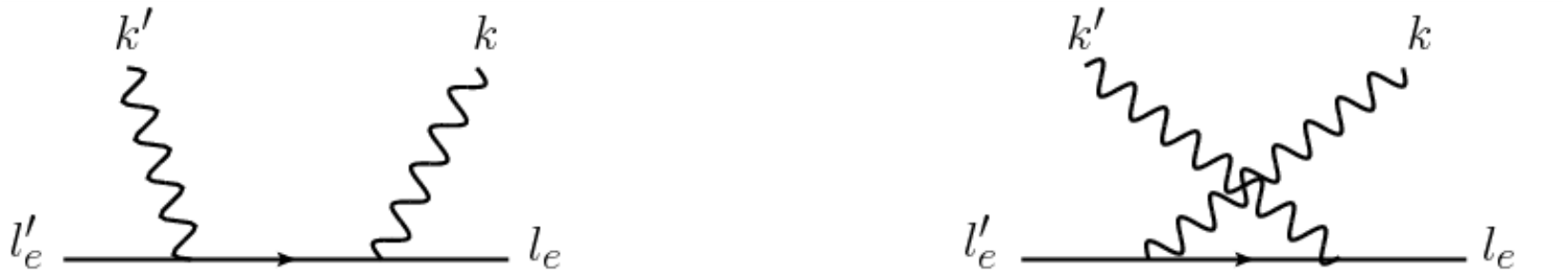}
\caption{Compton scattering diagrams which appear in ${\rm Im}({\cal M}_{\rm loop})$ for $\gamma-e$ cuts.
We denote the two graphs by ${\cal M}^{\rm d}_{\gamma e}(l'_e,k',l_e,k)$ 
and ${\cal M}^{\rm c}_{\gamma e}(l'_e,k',l_e,k)$, respectively. The diagrams and amplitudes appropriate
to $\gamma-p$ scattering follow from replacing electron with proton variables. 
}
\label{fig:compton}
\end{center}
\end{figure}
It is useful to categorize the cuts as per the sorts of processes involved. That is, 
${\cal M}_{fn}$ 
describes the manner in which select particles rescatter, so that we can have 
Compton scattering or electron-proton scattering, the latter with or without 
the emission of an additional photon. 
The family of diagrams given by (1), (2), (5.1), and (6.2) contains Compton scattering from the electron, as illustrated in 
Fig.~\ref{fig:compton}, whereas the family comprised of (3), (4), (7.2), and (8.3) contains Compton scattering from the proton. 
In these families ${\cal M}_{fn}$ is captured by one of the following expressions: 
\begin{equation}
{\cal M}^{\rm d}_{\gamma e}(l'_e,k',l_e,k)
=-e^2\bar{u}_e(l_e)\frac{2l_e\cdot\epsilon^\ast+\slashed{\epsilon}^\ast\slashed{k}}{2l_e\cdot k}\slashed{\epsilon}'u_e(l'_e) \,,
\end{equation}
\begin{equation}
{\cal M}^{\rm c}_{\gamma e}(l'_e,k',l_e,k)
=e^2\bar{u}_e(l_e)\frac{2l_e\cdot\epsilon'-\slashed{\epsilon}'\slashed{k'}}{2l'_e\cdot k}\slashed{\epsilon}^\ast u_e(l'_e) \,,
\end{equation}
\begin{equation}
{\cal M}^{\rm d}_{\gamma p}(p'_p,k',p_p,k)
=-e^2\bar{u}_p(p_p)\frac{2p_p\cdot\epsilon^\ast+\slashed{\epsilon}^\ast\slashed{k}}{2p_p\cdot k}\slashed{\epsilon}'u_p(p'_p) \,, 
\end{equation}
or
\begin{equation}
{\cal M}^{\rm c}_{\gamma p}(p'_p,k',p_p,k)
=e^2\bar{u}_p(p_p)\frac{2p_p\cdot\epsilon'-\slashed{\epsilon}'\slashed{k'}}{2p'_p\cdot k}\slashed{\epsilon}^\ast u_p(p'_p) \,,
\end{equation}
where $\epsilon'\equiv \epsilon(k')$. 
Correspondingly, ${\cal M}_{ni}$ is given by the tree-level neutron radiative 
$\beta$-decay amplitude, as per the form of ${\cal M}_{01}$ and 
${\cal M}_{02}$, with only some of the arguments changed. 
Technically we define a ``family'' to be those contributions to the \ot-odd
correlation which cancel among
themselves to yield zero when we replace $\epsilon$ or $\epsilon^\ast$ by $k$ or 
$\epsilon'$ or $\epsilon'^\ast$ by $k'$ as per the Ward-Takahashi identities. 

Furthermore, 
there is an intermediate phase-space integral over the kinematically allowed phase space. 
For $\gamma-e$ scattering we have 
\begin{equation}
\int d\rho_{\gamma e}\equiv\int \frac{d^3{\bf l'}_e}{2E'_e}\frac{d^3{\bf k'}}{2\omega'}\delta^{(4)}(l'_e+k'-P_{0e}) \,,
\end{equation}
with $P_{0e}\equiv l_e +k$, whereas for $\gamma-p$ scattering we have
\begin{equation}
\int d\rho_{\gamma p}\equiv\int \frac{d^3{\bf p'}_p}{2E'_p}\frac{d^3{\bf k'}}{2\omega'}\delta^{(4)}(p'_p+k'-P_{0p}) \,,
\end{equation}
with $P_{0p}\equiv p_p+k$. Collecting the pieces, we have 
\begin{eqnarray}
&& {\rm Im}({\cal M}_{1})=\frac{1}{8\pi^2}\int d\rho_{\gamma e}\sum_{s_{\gamma e}}
{{\cal M}^d_{\gamma e}(l'_e,k',l_e,k){\cal M}_{01}(l'_e,k',p_p)} \,,  \\ 
&& {\rm Im}({\cal M}_{2})=\frac{1}{8\pi^2}\int d\rho_{\gamma e}\sum_{s_{\gamma e}}
{{\cal M}^c_{\gamma e}(l'_e,k',l_e,k){\cal M}_{01}(l'_e,k',p_p)} \,, \\
&& {\rm Im}({\cal M}_{5.1})=\frac{1}{8\pi^2}\int d\rho_{\gamma e}\sum_{s_{\gamma e}}
{{\cal M}^d_{\gamma e}(l'_e,k',l_e,k){\cal M}_{02}(l'_e,k',p_p)} \,, \\
&& {\rm Im}({\cal M}_{6.2})=\frac{1}{8\pi^2}\int d\rho_{\gamma e}\sum_{s_{\gamma e}}
{{\cal M}^c_{\gamma e}(l'_e,k',l_e,k) {\cal M}_{02}(l'_e,k',p_p)} \,,
\end{eqnarray}
for the ``$\gamma-e$'' cuts, and 
\begin{eqnarray}
&& {\rm Im}({\cal M}_{3})=\frac{1}{8\pi^2}\int d\rho_{\gamma p}\sum_{s_{\gamma p}}
{{\cal M}^d_{\gamma p}(p'_p,k',p_p,k){\cal M}_{02}(l_e,k',p'_p)} 
\,, \\
&& {\rm Im}({\cal M}_{4})=\frac{1}{8\pi^2}\int d\rho_{\gamma p}\sum_{s_{\gamma p}}
{{\cal M}^c_{\gamma p}(p'_p,k',p_p,k){\cal M}_{02}(l_e,k',p'_p)} 
\,, \\
&& {\rm Im}({\cal M}_{7.2})=\frac{1}{8\pi^2}\int d\rho_{\gamma p}\sum_{s_{\gamma p}}
{{\cal M}^d_{\gamma p}(p'_p,k',p_p,k){\cal M}_{01}(l_e,k',p'_p)} 
\,, \\
&& {\rm Im}({\cal M}_{8.3})=\frac{1}{8\pi^2}\int d\rho_{\gamma p}\sum_{s_{\gamma p}}
{{\cal M}^c_{\gamma p}(p'_p,k',p_p,k){\cal M}_{01}(l_e,k',p'_p)} \,,
\end{eqnarray}
for the $\gamma-p$ cuts.

\begin{figure}
\begin{center}
\includegraphics[scale=0.4]{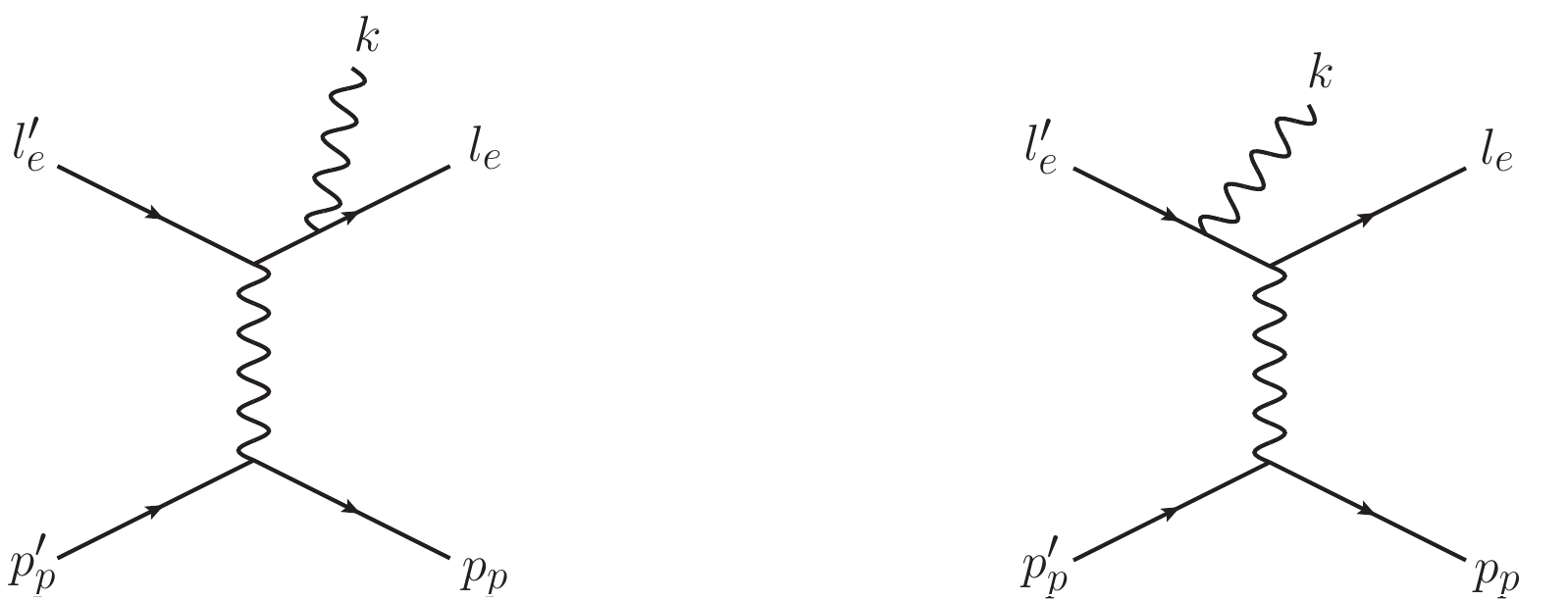}
\caption{Diagrams which appear in ${\rm Im}({\cal M}_{\rm loop})$ for $e-p$ scattering with electron bremsstrahlung. 
We denote the two graphs by ${\cal M}^{\rm ef}_{ep\gamma}(l'_e,p'_p,l_e,k,p_p)$ and 
${\cal M}^{\rm ei}_{ep\gamma}(l'_e,p'_p,l_e,k,p_p)$, respectively. 
The diagrams and amplitudes appropriate to 
proton bremsstrahlung 
follow from exchanging electron and proton variables. 
}
\label{fig:epgamma}
\end{center}
\end{figure}

\begin{figure}
\begin{center}
\includegraphics[scale=0.5]{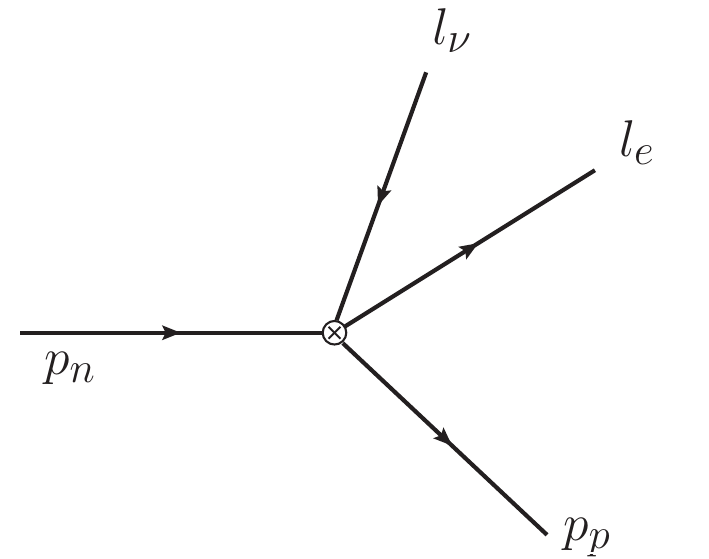}
\caption{Contribution to $n(p_n)\to p(p_p) + e^-(l_e) + {\bar\nu}_e(l_\nu)$ decay after 
Fig.~\ref{fig:tree}.}
\label{fig:DK}
\end{center}
\end{figure}
In addition to the families of Compton cuts, there are cuts in which 
${\cal M}_{fn}$ is determined by electron-proton scattering either with and without 
bremsstrahlung, and, correspondingly, ${\cal M}_{ni}$ is determined by either 
nonradiative or radiative $\beta$ decay. Referring to Fig.~\ref{fig:allcuts}, we see 
for cuts in which the electron and proton scatter with bremsstrahlung 
that diagrams 
(5.2) and (6.1) comprise the family associated with electron bremsstrahlung, 
as shown in Fig.~\ref{fig:epgamma}, and 
(7.1) and (8.1) comprise the family associated with proton bremsstrahlung. 
In these families ${\cal M}_{fn}$ is given by one of the following: 
\begin{equation}
{\cal M}^{\rm ef}_{ep\gamma}(l'_e,p'_p,l_e,k,p_p)=-e^3\bar{u}_e(l_e)
\frac{2l_e\cdot\epsilon^\ast+\slashed{\epsilon}^\ast\slashed{k}}{2l_e\cdot k}\gamma^\mu u_e(l'_e)
\frac{g_{\mu\nu}}{(p'_p-p_p)^2}\bar{u}_p(p_p)\gamma^\nu u_p(p'_p) \,,
\end{equation}
\begin{equation}
{\cal M}^{\rm ei}_{ep\gamma}(l'_e,p'_p,l_e,k,p_p)=e^3\bar{u}_e(l_e)\gamma^\mu\frac{2l'_e\cdot\epsilon^\ast-\slashed{k}
\slashed{\epsilon}^\ast}{2l'_e\cdot k} u_e(l'_e)\frac{g_{\mu\nu}}{(p'_p-p_p)^2}\bar{u}_p(p_p)\gamma^\nu u_p(p'_p) \,,
\end{equation}
\begin{equation}
{\cal M}^{\rm pf}_{ep\gamma}(l'_e,p'_p,l_e,k,p_p)=e^3\bar{u}_p(p_p)\frac{2p_p\cdot\epsilon^\ast+\slashed{\epsilon}^\ast
\slashed{k}}{2p_p\cdot k}\gamma^\mu u_p(p'_p)\frac{g_{\mu\nu}}{(l'_e-l_e)^2}\bar{u}_e(l_e)\gamma^\nu u_e(l'_e) \,,
\end{equation}
or 
\begin{equation}
{\cal M}^{\rm pi}_{ep\gamma}(l'_e,p'_p,l_e,k,p_p)=-e^3\bar{u}_p(p_p)\gamma^\mu\frac{2p'_p\cdot\epsilon^\ast-
\slashed{k}\slashed{\epsilon}^\ast}{2p'_p\cdot k} u_p(p'_p)\frac{g_{\mu\nu}}{(l'_e-l_e)^2}\bar{u}_e(l_e)\gamma^\nu u_e(l'_e) \,.
\end{equation}
Moreover, 
${\cal M}_{ni}$ is given by neutron $\beta$ decay, as shown in Fig.~\ref{fig:DK}, which, up
to recoil-order corrections, reads:
\begin{equation}
{\cal M}_{\rm DK}(l'_e,p'_p)=\frac{g_VG_F}{\sqrt{2}}\bar{u}_e(l'_e)\gamma_\rho(1-\gamma_5)v_\nu(l_\nu)\bar{u}_p(p'_p)\gamma^\rho(1-\lambda\gamma_5)u_n(p_n) \,.
\end{equation}
Collecting the pieces, we have 
\begin{eqnarray}
&& {\rm Im}({\cal M}_{5.2})=\frac{1}{8\pi^2}\int d\rho_{ep\gamma} \sum_{s_{ep}}
{{\cal M}^{\rm ef}_{ep\gamma}(l'_e,p'_p,l_e,k,p_p) {\cal M}_{\rm DK}(l'_e,p'_p)} \,,\\
&& {\rm Im}({\cal M}_{6.1})=\frac{1}{8\pi^2}\int d\rho_{ep\gamma} \sum_{s_{ep}}
{{\cal M}^{\rm ei}_{ep\gamma}(l'_e,p'_p,l_e,k,p_p) {\cal M}_{\rm DK}(l'_e,p'_p)} \,,
\end{eqnarray}
and 
\begin{eqnarray}
&& {\rm Im}({\cal M}_{7.1})=\frac{1}{8\pi^2}\int d\rho_{ep\gamma} \sum_{s_{ep}} 
{{\cal M}^{\rm pf}_{ep\gamma}(l'_e,p'_p,l_e,k,p_p){\cal M}_{\rm DK}(l'_e,p'_p)} \,, \\
&& {\rm Im}({\cal M}_{8.1})=\frac{1}{8\pi^2}\int d\rho_{ep\gamma} \sum_{s_{ep}}
{{\cal M}^{\rm pi}_{ep\gamma}(l'_e,p'_p,l_e,k,p_p) {\cal M}_{\rm DK}(l'_e,p'_p)} \,
\end{eqnarray}
for the ``$e-p-\gamma$'' cuts. 
The last family of cuts is given by (6.3) and (8.2) in Fig.~\ref{fig:allcuts}. In this case 
${\cal M}_{fn}$ is given by $e-p$ scattering, and we have
\begin{equation}
{\cal M}_{ep}(l'_e,p'_p,l_e,p_p)
=-e^2\bar{u}_e(l_e)\gamma^\mu u_e(l'_e)\frac{g_{\mu\nu}}{(l'_e-l_e)^2}
\bar{u}_p(p_p)\gamma^\nu u_p(p'_p) \,. 
\end{equation}
The corresponding ${\cal M}_{ni}$ is given by ${\cal M}_{01}(l'_e,k,p'_p)$ for (6.3) 
and ${\cal M}_{02}(l'_e,k,p'_p)$ for (8.2). We thus have
\begin{eqnarray}
&& {\rm Im}({\cal M}_{6.3})=\frac{1}{8\pi^2}\int d\rho_{ep} \sum_{s_{ep}} 
{{\cal M}_{ep}(l'_e,p'_p,l_e,p_p){\cal M}_{01}(l'_e,k,p'_p)} \,, \\
&& {\rm Im}({\cal M}_{8.2})=\frac{1}{8\pi^2}\int d\rho_{ep} \sum_{s_{ep}}
{{\cal M}_{ep}(l'_e,p'_p,l_e,p_p){\cal M}_{02}(l'_e,k,p'_p)} \,
\end{eqnarray}
for the $e-p$ cuts. 
In these graphs the intermediate momenta satisfy $l'_e + p'_p = l_e + p_p$, so that the integral over the allowed
phase space is slightly different from that in families with $e-p$ scattering and bremsstrahlung. In particular, 
diagrams (6.3) and (8.2) are each infrared divergent when $l_e'=l_e$; this divergence cancels, however, as expected~\cite{KLN}, 
once we construct $A_\xi^{\rm SM}$. 

The expressions we have collected complete the building blocks of the computation 
of the \ot-odd correlation in 
${\cal O}(\alpha)$ up to recoil-order corrections. The spin-averaged 
\ot-odd correlation is
\begin{equation}
{\overline{|{\cal M}|^2}_{\rm \ot \ odd}} \equiv
\frac{1}{2} \sum_{\rm spins} |{\cal M}|^2_{\rm \ot \ odd} 
= \frac{1}{2} 
\sum_{\rm spins} (2 {\rm Re}({\cal M}_{\rm tree} i {\rm Im} {\cal M}^\ast_{\rm loop})) \,,
\end{equation}
and we report the contributions to it family by family as each family represents
a QED gauge-invariant group of contributions. 
We employ a subscript system to identify the contributions 
in a straightforward way. Since the \ot-odd correlations are given by 
the interference of tree-level diagrams, which are numbered in 
Fig.~\ref{fig:tree} as (01) and (02), with one-loop level diagrams, 
which are numbered in Fig.~\ref{fig:allcuts} as (1), (2),..., and (8.3), 
we label, for example, the \ot-odd correlation from the tree diagram (01) and 
one-loop diagram (6.3) as ${\overline{|{\cal M}|^2}_{\rm \ot \ odd}}[6.3.01]$. 
In computing the intermediate phase-space integrals which enter these expresssions, 
we find that both vector and tensor structures appear in the intermediate momenta. 
We simplify such integrals using the Passarino-Veltman reduction~\cite{passvelt} 
and present the details, as well as all needed integrals, in Appendix A. 
In Appendix B we report concrete expressions for the final gauge-invariant 
combinations of the various contributions 
to ${\overline{|{\cal M}|^2}_{\rm \ot \ odd}}$ 
which result after performing the trace calculations and employing
the formulas of Appendix A for the intermediate phase-space integrals. 
We work to leading order 
in the recoil expansion throughout. 
Judging by the structure of the resulting expressions 
one can see that some families, namely, the 
$\gamma-p$ family containing cuts (3)+(4)+(7.2)+(8.3), 
as well as the $e-p-\gamma$ family containing cuts (7.1)+(8.1), 
do not have leading-recoil-order contributions, whereas 
others do and need to be considered carefully. 
The computations necessary 
%detailed computations 
to determine ${\rm Im}({\cal M}_{\rm loop})$ and the resulting \ot-odd 
interference term are involved, 
so that we employ the program FORM to compute analytic 
expressions for the traces~\cite{vermaseren}. 
We compute all of the diagrams with these methods as a check of our procedures
-- we verify that the expected cancellations do indeed occur.

\section{Results}
\label{res}

Before presenting our final results for the asymmetry, there are three important remarks 
to be made concerning our numerical evaluation of the integral of 
${\overline{|{\cal M}|^2}_{\rm \ot \ odd}}$ over the allowed phase space. 
First of all, we note that 
the contributions to 
the asymmetry from the $e-p$ and $e-p-\gamma$ cuts dominate the final numerical 
result. The $\gamma-p$ contribution vanishes in leading order, whereas the 
$\gamma-e$ contribution partially cancels -- the latter observation 
comes from our detailed numerical 
evaluation of the asymmetry. 
Second, we 
note that the contributions from the diagrams of the 
$e-p$ cuts 
each contain an infrared divergence; we regulate this by 
inserting a fictitious photon mass $m_\gamma$. 
However, as we show in 
Appendix B, 
the infrared divergence cancels in 
the net 
contribution to the asymmetry from the $e-p$ cuts. 
The remaining piece is thus finite and well-defined, and 
we can safely set $m_\gamma$ to zero.  
Finally, we note it is most convenient to choose a 
restricted range in the $\gamma-e$ opening angle. 
As one can see from the formulas in the Appendix A, 
the solutions to the Passarino-Veltman equations 
become invalid if the opening angle $\theta_{e\gamma}$ 
between the outgoing electron and the photon is exactly 
equal to 0 or to $\pi$. There is no physical divergence. 
Rather, the spatial components 
of the vector and tensor equations to determine 
the relevant coefficients become degenerate at such a boundary. 
Potentially one could remove this difficulty by solving
the equations for infinitesimal values of $\theta_{e\gamma}$ or 
$(\theta_{e\gamma}-\pi)$ 
and then 
interpolating the solutions to the needed $\theta_{e\gamma}=0$ and $\pi$ points. 
In our present work, we 
simply choose a restricted 
range $x_k\equiv \cos \theta_{e\gamma}\in [-0.9,0.9]$, which spans the 
angular range over which the neutron radiative decay rate is largest~\cite{byrne}.

\begin{table}[t]   
\caption{\label{nrdk}  
\ot-odd asymmetry as a function of $\omega_{\rm min}$ for 
neutron radiative $\beta$ decay. 
} \centering   \footnotesize
\vskip 0.5\baselineskip
\begin{center}
\begin{tabular}{l*{6}{c}r}
$\omega_{\rm min}({\rm MeV})$ && ${A_\xi}$ \\
\hline
0.01 && $1.76\times10^{-5}$  \\
0.05 && $3.86\times10^{-5}$  \\
0.1 && $6.07\times10^{-5}$  \\
0.2 && $9.94\times10^{-5}$  \\
0.3 && $1.31\times10^{-4}$ \\
0.4 && $1.54\times10^{-4}$ \\
0.5 && $1.70\times10^{-4}$ \\
0.6 && $1.81\times10^{-4}$ \\
0.7 && $1.89\times10^{-4}$ \\
\hline
\end{tabular}
\end{center}
\vspace{1ex} 
\end{table}

\begin{figure}
\begin{center}
\includegraphics[scale=0.6]{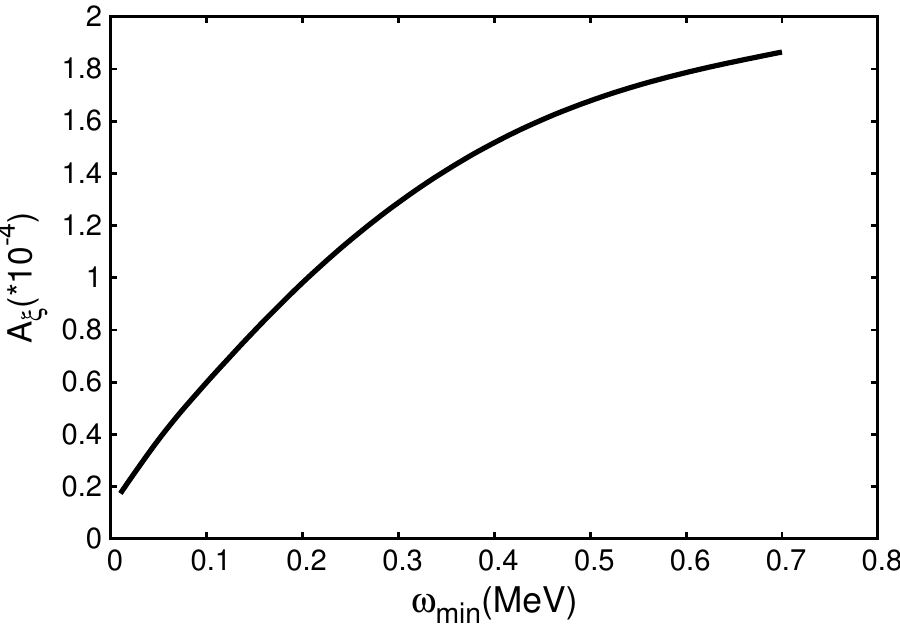}
\caption{The asymmetry $A_{\xi}$ versus the smallest 
detectable photon energy $\omega_{\rm min}$ in neutron radiative $\beta$ decay.}
\label{fig:AvsEkmin}
\end{center}
\end{figure}

We can now present our results for $A_\xi^{\rm SM}$. Noting Eq.~(\ref{asym}), 
we see that Eqs.~(\ref{calM0sq}), and  (\ref{tensors}) 
share a common factor of $e^2 g_V^2 G_F^2 M^2/2$, 
making $A_\xi^{\rm SM}$ independent of the decaying particle's mass
in leading order in the recoil expansion. 
As can be seen explicitly in Appendix B, 
all of the contributions to ${\overline{|{\cal M}|^2}_{\rm \ot \ odd}}$ 
are found to be proportional to $(1-\lambda^2)$, 
so that the resulting asymmetry goes 
as $(1-\lambda^2)/(1+3\lambda^2)$, up to small corrections, in this limit. 
The dependence on $\lambda$ in ${\overline{|{\cal M}|^2}_{\rm \ot \ odd}}$ 
stems from the special nature of the 
\ot-odd correlation. It is a real triple product in momenta 
arising from the interference of a tree-level diagram 
with an imaginary part of a one-loop diagram after summing over the particles' spins. 
To leading order in $M$, the 
only surviving contribution is obtained from 
the product of the symmetric part of the lepton tensor, which is determined by a 
trace containing $\gamma_5$, namely, 
$l_\nu^\rho \epsilon^{\alpha\beta\gamma\delta}
+l_\nu^\delta \epsilon^{\alpha\beta\gamma\rho}-g^{\rho\delta}l_{\nu\,\mu} 
\epsilon^{\alpha\beta\gamma\mu}$, 
where $\alpha, \beta$, and $\gamma$ refer to photon or lepton indices, 
with the symmetric part of the hadron tensor. The latter is proportional to 
$(1+\lambda^2)p_\rho p_\delta-\lambda^2 M^2 g_{\rho\delta}$, 
where $p$ is a baryon momentum and $p^2=M^2$. 
As one can easily check, 
this special combination generates an overall $(1-\lambda^2)$ coefficient; the
remaining $(1+\lambda^2)$ term cannot be of leading order once the photon spin
sum is effected. We use $\lambda =  -1.2701 \pm 0.0025$~\cite{pdg} in our numerical
evaluation. For definiteness, the remaining 
%input parameters we employ are $m_e=0.510998910\,\hbox{MeV}$, 
%$M_n=939.565346\,\hbox{MeV}$, $M_p=938.272013\,\hbox{MeV}$, and
%$\alpha^{-1} =137.035999679$ -- these quantities 
input parameters we employ are $m_e=0.510999\,\hbox{MeV}$, 
$M_n=939.565\,\hbox{MeV}$, $M_p=938.272\,\hbox{MeV}$, and
$\alpha^{-1} =137.0360$ -- these quantities 
can be regarded as exact for our current purpose~\cite{pdg}. 
We show our results for the \ot-odd asymmetry in neutron radiative $\beta$ decay in 
Table \ref{nrdk} and Fig.~\ref{fig:AvsEkmin}. 
We see that the asymmetry is rather
smaller than $\alpha$. 
We recall that the radiative $\beta$-decay
rate grows as $\log \omega_{\rm min}$ as $\omega_{\rm min}\to 0$, whereas the 
${\overline{|{\cal M}|^2}_{\rm \ot \ odd}}$ 
tends to zero in that limit. Consequently the small values of the
asymmetry as $\omega_{\rm min}\to 0$ is reflective by the growth in 
the decay rate itself. 

Generally $A_\xi^{\rm SM}$ is determined by an interplay between 
$\lambda$ and the energetics
of the decay, along with the value of $\omega_{\rm min}$. 
The $(1-\lambda^2)$ behavior of ${\overline{|{\cal M}|^2}_{\rm \ot \ odd}}$ 
we have found in neutron radiative $\beta$ decay, neglecting terms of recoil order, 
is universal to allowed nuclear radiative $\beta$ decay in this limit as well. 
In the case of the decay of a $J=1/2$ nucleus this follows because we can treat the parent 
and daughter nuclei as elementary fermions while evaluating the electromagnetic radiative corrections. 
For the decay of a nucleus of arbitrary $J$, the result follows from the use of the impulse approximation 
for a $\beta$ decay at tree level. 
The $(1-\lambda^2)/(1+3\lambda^2)$ behavior of  $A_\xi^{\rm SM}$ in $\lambda$ makes for a rich pattern. 
If, for some nucleus, the associated value of $\lambda$ 
were significantly different from unity, 
the \ot-odd effect could be considerably amplified, 
whereas for if $\lambda\sim 1$, 
the \ot-odd effect could be substantially reduced, facilitating from this
perspective at least the search for physics beyond the SM. Interestingly 
a ``quenching'' of the Gamow-Teller strength 
in nuclei in relation to shell-model predictions is experimentally
established~\cite{Wildenthal:1983zz,caurier} -- 
it derives from the presence of many-body correlations in the nucleus~\cite{quench}. 
As a concrete example,  we consider the 
process $^{19}{\rm Ne}\rightarrow {^{19}{\rm F}}+e^+ +\nu_e+\gamma$. 
The $^{19}{\rm Ne}$ lifetime is much shorter than that of the neutron, making experiments more 
practical, and it should be possible to study such decays in a trapped atom experiment~\cite{shimizu}. 
Moreover, in this decay the axial-vector coupling is given by $g_A^{\rm eff}=0.928$, as determined 
by Refs.~\cite{babrown,brownWild} with Ref.~\cite{holstein} for a translation from the conventions of those references to $g_A^{\rm eff}$. Conseguently, 
we expect the asymmetry in $^{19}{\rm Ne}$ radiative $\beta$ decay to be smaller than that in the 
neutron case; we reserve detailed numerical results, however for a subsequent paper~\cite{svgdh2}. 

In our paper, we compute the ${\cal O}(\alpha)$ contribution to the \ot-odd asymmetry, 
keeping only the leading terms in the recoil expansion. The accuracy of our calculation
is limited by the uncertainties in 
the input parameters we employ, as well as by the numerical 
size of the neglected recoil-order contributions. Crudely we expect the latter
to be reduced with respect to the leading-order contribution 
by a factor of ${\cal O}(E_e^{\rm max}/M)\sim 1\times 10^{-3}$. 
%The uncertainty in $\lambda^{\rm eff}$ may well be
%the dominant error in some nuclei. 
Nevertheless, we can conveniently check 
the rough size of the recoil-order contributions in the neutron 
case by replacing the 
vertex $\gamma^\mu(g_V- g_A\gamma_5)$ in the tree-level amplitude 
with the weak magnetism contribution, 
$-i\sigma^{\mu\nu}q_\nu F_2(q^2)/(2M)$, where $F_2(0)/g_V=\kappa_v=3.706$, the 
isovector magnetic moment of the nucleon. 
The interference of the resulting recoil-order contribution 
with the tree-level amplitude yields upon 
explicit calculation a contribution to $A_\xi^{\rm SM}$ 
which is no larger than $\sim 4\times 10^{-7}$ 
for $\omega_{\rm min}= 0.3\,{\rm  MeV}$.

\section{Summary} 

In this paper, we have computed the \ot-odd correlation in neutron 
radiative $\beta$ decay arising from SM physics. 
The \ot-odd correlation is 
characterised by the kinematical variable 
$\xi = {\mathbf l}_\nu\cdot({\mathbf l}_e\times{\mathbf k})$; consequently, 
it is spin independent -- and thus fundamentally different from 
a permanent EDM. 
The mimicking \ot-odd correlation arises from the presence of electromagnetic
final-state interactions when the intermediate 
particles are each put on their own mass shell. 
%Taking the neutron case as a prototype,
We have computed the leading-order result, which is of ${\cal O}(\alpha,(\varepsilon/M)^0)$,
to the \ot-odd asymmetry exactly. 
In particular, our detailed analysis shows that
the resulting \ot-odd asymmetry is controlled by $(1-\lambda^2)$, so that 
$A_\xi^{\rm SM}$ vanishes as $\lambda\rightarrow 1$, suggesting that 
% hyperons
radiative $\beta$-decay studies in other systems could be employed to good effect. 
We will report our computation of the \ot-odd correlation in nuclear radiative $\beta$-decay
in a subsequent paper; there are additional Feynman diagrams, but they, up to 
corrections of recoil order, cancel to 
yield the gauge-invariant combinations of graphs we have computed in this paper~\cite{svgdh2}. 
%Indeed our procedure is completely general, so that  the resulting formulae 
%can be readily translated to hyperon and nuclear radiative $\beta$ decays as well. 
%We have explicitly considered the example of $^{19}{\rm Ne}$ radiative $\beta$ decay 
% as a concrete illustration of the possibilities. 
%Its lifetime is much shorter than the neutron's, making experiments more practical, 
%and it should be possible to study such decays in a trapped 
%atom experiment~\cite{shimizu}. 
%It also possesses a value of $\lambda^{\rm eff}$ rather close to unity. 
%The combined 
%effect of these observations is a \ot-odd
%asymmetry roughly an order of magnitude smaller than that in neutron $\beta$ decay. 
%In this case $A_\xi^{\rm SM}$ presents a much reduced background against
%which to search for new sources of  \cp-violation. 

\acknowledgments
We are grateful to J. Vermaseren and the FORM forum for helpful assistance
in the use of FORM. We thank T. Gentile for information regarding
the $\gamma-e$ opening angle dependence of the neutron radiative $\beta$-decay rate, and 
S.G. thanks the Aspen Center for Physics for hospitality 
during the execution of this work. We acknowledge partial support from the 
U.S. Department of Energy under contract DE-FG02-96ER40989. 

\appendix

\section{Intermediate Phase-Space Integrals}

The computation of the imaginary parts of the loop diagrams 
requires an integration over the allowed phase space of the intermediate momenta as fixed 
by the momenta of the final-state particles and energy-momentum conservation. 
In this Appendix 
we report the integrals which appear in the diagrams of Fig.~\ref{fig:allcuts} and 
label them as per the diagrams in that figure. 
For diagrams with cuts which yield Compton scattering from electrons
our results can be compared to, and agree with, those of Refs.~\cite{braguta,khriprud}. 
In what follows we report the integrals which
arise from $\gamma-e$ cuts: (1), (2), (5.1), and (6.2), 
and then the integrals which 
arise from the cutting of electron
and proton lines 
to generate physical $ep\to ep\gamma$ scattering, namely, (5.2) and (6.1), 
and $ep\to ep$
scattering, (6.3) and (8.2). The integrals 
associated with the remaining cuts in Fig.~\ref{fig:allcuts} 
are not given explicitly because they do not contribute 
in leading order in the recoil expansion, as we 
note in the main body of the text. 
Nevertheless, we note the relationships between these integrals which appear 
in the large $M_p$ limit in order to make the cancellations associated 
with these terms transparent. 

From diagram $(1)$, defining $P_{0e}\equiv l_e+ k$,  we have 
\begin{eqnarray}
J_1\equiv \int \frac{d^3\mathbf{l}_e^\prime}{2E_e^\prime}\frac{d^3 \mathbf{k}^\prime}{2\omega^\prime}
\delta^{(4)}(l_e'+ k'- P_{0e}) \nonumber &\equiv& \int d\rho_{\gamma e} \nonumber \\
&=& \frac{\pi}{2}\left(1-\frac{m_e^2}{P^2_{0e}}\right)\,,
\end{eqnarray}
as well as 
\begin{equation}
%\int \frac{d^3{\bf Le'}}{2Ee'}\frac{d^3{\bf K'}}{2\omega'}K'^\mu\delta^{(4)}(Le'+K'-P_0)=a_PP^\mu_0
K_1^\mu\equiv \int d\rho_{\gamma e} k^{\prime\,\mu}=a_1P_{0e}^\mu
\end{equation}
with
\begin{equation}
\nonumber a_1=\frac{\pi}{4}\left(1-\frac{m_e^2}{P_{0e}^2}\right)^2 \,.
\end{equation}
From diagram $(2)$ we have 
\begin{equation}
J_2 \equiv \int d\rho_{\gamma e} \frac{1}{l_e\cdot k^\prime} = 
%\int \frac{d^3{\bf Le'}}{2Ee'}\frac{d^3{\bf K'}}{2\omega'}\frac{1}{Le'\cdot K}\delta^{(4)}(Le'+K'-P_0)=b_0=
\frac{\pi}{2l_e\cdot k}\log\left(\frac{P_{0e}^2}{m_e^2}\right) \,.
\end{equation}
We apply the Passarino-Veltman reduction method to compute integrals which contain 
additional powers of the intermediate momenta~\cite{passvelt}. 
That is, writing 
\begin{equation}
K_2^\mu = \int d\rho_{\gamma e} \frac{k^{\prime\,\mu}}{l_e \cdot k^\prime} = a_2 l_e^\mu+ b_2 P_{0e}^\mu \,,
\end{equation}
the values of $a_2$ and $b_2$ are fixed by the solution of 
the set of equations
\begin{eqnarray}
\nonumber J_1 &=& a_2 m_e^2+ b_2 l_e\cdot P_{0e} \,,\\
\nonumber l_e\cdot k J_2 &=& a_2 l_e\cdot P_{0e} + b_2 P_{0e}^2 \,.
\end{eqnarray}
Moreover, 
\begin{equation}
L_2^{\mu \nu} = \int d\rho_{\gamma e} \frac{k^{\prime \,\mu}k^{\prime \,\nu}}{l_e\cdot k^\prime}
=c_2g^{\mu\nu}+d_2 l_e^\mu l_e^\nu + e_2 P_{0e}^\mu P_{0e}^\nu + f_2(l_e^\mu P_{0e}^\nu +P_{0e}^\mu l_e^\nu)\,,
\end{equation}
where $c_2$, $d_2$, $e_2$, and $f_2$ are given by the solution of the set of equations
\begin{eqnarray}
\nonumber 0 &=& 4c_2+d_2 m_e^2+ e_2 P_{0e}^2+  2 f_2l_e\cdot P_{0e} \,,\\
\nonumber 0 &=& c_2+ d_2 m_e^2+ f_2 l_e \cdot P_{0e} \,, \\
\nonumber a_1 &=& e_2 l_e\cdot P_{0e} + f_2 m_e^2 \,,\\
\nonumber l_e \cdot k b_2 &=& c_2+ e_2 P_{0e}^2 + f_2 l_e\cdot P_{0e} \,.
\end{eqnarray}

For integrals which depend on $M_p$ we report their form in the large $M_p$ limit 
for subsequent use. Note that $M$ rather
than $M_p$ appears in the limiting form because 
the $n-p$ mass difference itself is of higher order in the recoil expansion. 
From diagram $(5.1)$ we have 
\begin{equation}
J_{5.1} = \int d\rho_{\gamma e} \frac{1}{p_p\cdot k'} = \frac{\pi}{2 I_{0e}} \log \left( 
\frac{p_p \cdot P_{0e} + I_{0e}}{p_p \cdot P_{0e} - I_{0e}}
\right) \,,
\end{equation}
with $I_{0e} = \sqrt{(p_p\cdot P_{0e})^2 - M_p^2 P_{0e}^2}$, noting 
\begin{equation}
J_{5.1} \sim \frac{\pi}{2M |\mathbf{k} + \mathbf{l_e}|}
\log\left( \frac{E_e + \omega + |\mathbf{k} + \mathbf{l_e}|}{E_e + \omega - |\mathbf{k} + \mathbf{l_e}|}
 \right) \,
\end{equation}
as ${M_p\to \infty}$. In addition
\begin{equation}
K_{5.1}^\mu = \int d\rho_{\gamma e} \frac{k'^\mu}{p_p \cdot k'} = a_{5.1} p_p^\mu + b_{5.1} P_{0e}^\mu \,,
\end{equation}
where $a_{5.1}$ and $b_{5.1}$ are given by the solution of the set of equations
\begin{eqnarray}
\nonumber J_1 &=& a_{5.1} M_p^2 + b_{5.1} p_p\cdot P_{0e} \,, \\
\nonumber l_e\cdot k J_{5.1} &=& a_{5.1} p_p\cdot P_{0e} + b_{5.1} P_{0e}^2 \,.
\end{eqnarray}
In the large $M_p$ limit $b_{5.1} \sim 1/M$ and $a_{5.1} \sim 1/M^2$. 
We postpone discussion of the integrals from diagrams $(5.2)$ and $(6.1)$ to consider
the integrals from the remaining diagrams with Compton cuts. 
From diagram $(6.2)$ we have 
% also 8.3
\begin{equation}
J_{6.2}=\int d\rho_{\gamma e} \frac{1}{(l_e\cdot k')(p_p\cdot k')}
=\frac{\pi}{2(l_e\cdot k)I_e} \log\left(\frac{p_p\cdot l_e + I_e}{p_p\cdot l_e - I_e}\right) \,,\\
\end{equation}
with $I_e=\sqrt{(p_p\cdot l_e)^2-M_p^2m_e^2}$ and 
\begin{equation}
J_{6.2} \sim \frac{\pi}{2 M |\mathbf{l}_e| k\cdot l_e} \log\left(
\frac{E_e + |\mathbf{l}_e|}{E_e - |\mathbf{l}_e|} 
\right) \,
\end{equation}
as ${M_p\to \infty}$. In addition
\begin{equation}
K_{6.2}^\mu = \int d\rho_{\gamma e} \frac{k'^\mu}{(l_e\cdot k')(p_p\cdot k')}
=a_{6.2} P_{0e}^\mu + b_{6.2} l_e^\mu+c_{6.2} p_p^\mu \,,
\end{equation}
where $a_{6.2}$, $b_{6.2}$, and $c_{6.2}$ are given by the solution to the set of equations
\begin{eqnarray}
\nonumber J_2 &=& a_{6.2}p_p\cdot P_{0e}+b_{6.2}p_p\cdot l_e+c_{6.2}M_p^2 \,,\\
\nonumber J_{5.1} &=& a_{6.2} l_e\cdot P_{0e} +b_{6.2} m_e^2+c_{6.2} p_p\cdot l_e \,, \\
\nonumber l_e \cdot k J_{6.2} &=&   a_{6.2} P_{0e}^2+b_{6.2} l_e\cdot P_{0e} +c_{6.2} p_p\cdot P_{0e} \,,
\end{eqnarray}
%and in the large $M_p$ limit $a_{6.2}, b_{6.2} \sim 1/M_p$ and $c_{6.2} \sim 1/M_p^2$. Finally
and in the large $M_p$ limit $a_{6.2}, b_{6.2} \sim 1/M$ and $c_{6.2} \sim 1/M^2$. Finally
\begin{eqnarray}
\nonumber L_{6.2} ^{\mu\nu} &=& \int d\rho_{\gamma e} \frac{k'^\mu k'^\nu}{(l_e\cdot k')(p_p\cdot k')} \\
\nonumber &=& d_{6.2}g^{\mu\nu}+e_{6.2}p_p^\mu p_p^\nu+f_{6.2}l_e^\mu l_e^\nu +g_{6.2}P_{0e}^\mu P_{0e}^\nu + 
h_{6.2}(p_p^\mu l_e^\nu+ l_e^\mu p_p^\nu) \\
 && +i_{6.2}(p_p^\mu P_{0e}^\nu+ P_{0e}^\mu p_p^\nu)
+k_{6.2}(l_e^\mu P_{0e}^\nu + P_{0e}^\mu l_e^\nu) \,,
\end{eqnarray}
where the coefficients which appear are given by the solution to 
set of the equations
\begin{eqnarray}
\nonumber && 4d_{6.2}+e_{6.2}M_p^2+f_{6.2}m_e^2+g_{6.2}P_{0e}^2+2h_{6.2} p_p\cdot l_e+2i_{6.2}p_p\cdot P_{0e}
+2k_{6.2} l_e\cdot P_{0e}=0 \,,\\
\nonumber && d_{6.2}+f_{6.2} m_e^2+h_{6.2}p_p\cdot l_e +k_{6.2} l_e\cdot P_{0e} =0 \,,\\
\nonumber && e_{6.2} p_p\cdot l_e+h_{6.2} m_e^2+i_{6.2} l_e\cdot P_{0e}=a_{5.1} \,,\\
\nonumber && g_{6.2} l_e\cdot P_{0e} +i_{6.2}p_p\cdot l_e +k_{6.2} m_e^2=b_{5.1} \,,\\
\nonumber && d_{6.2}+e_{6.2}M_p^2+h_{6.2}p_p\cdot  l_e+i_{6.2} p_p\cdot P_{0e}=0 \,,\\
\nonumber && g_{6.2} p_p\cdot P_{0e} +i_{6.2}M_p^2+k_{6.2}p_p\cdot l_e =b_2 \,,\\
\nonumber && d_{6.2}+g_{6.2}P_{0e}^2+i_{6.2}p_p\cdot P_{0e}+k_{6.2}l_e\cdot P_{0e}=l_e\cdot k a_{6.2}\,.
\end{eqnarray}
Note that the equations have been chosen to yield a self-consistent solution for the six coefficients. 

The integrals associated with the $\gamma-p$ cuts 
can be found if necessary 
by replacing the intermediate momentum $l_e'$ by $p_p'$ as well 
as $l_e$ by $p_p$ 
in the $\gamma-e$ integrals we have provided. Specifically we note 
\begin{eqnarray}
J_3\equiv \int \frac{d^3\mathbf{p}_p^\prime}{2E_p^\prime}\frac{d^3 \mathbf{k}^\prime}{2\omega^\prime}
\delta^{(4)}(p_p'+ k'- P_{0p}) \nonumber &\equiv& \int d\rho_{\gamma p} \,,
%&=& \frac{\pi}{2}\left(1-\frac{M_p^2}{P^2_{0p}}\right)\,,
\end{eqnarray}
where $P_{0p} \equiv p_p + k$, and 
\begin{equation}
J_4 = \int d\rho_{\gamma p} \frac{1}{p_p \cdot k'} 
%= \frac{\pi}{2 p_p\cdot k} \log\left( \frac{P_{0p}^2}{M_p^2} \,,
%\right) 
\end{equation}
so that 
\begin{equation}
J_4 \sim \frac{1}{M\omega} J_3 \sim  {\cal O}\left(\frac{1}{M^2}\right)
\label{asym34}
\end{equation}
as $M_p\to \infty$. 
Moreover, 
\begin{equation}
J_{7.2} = \int d\rho_{\gamma p} \frac{1}{l_e\cdot k'} 
\end{equation}
and 
\begin{equation}
K_{7.2}^\mu = \int d\rho_{\gamma p} \frac{k'^\mu}{l_e\cdot k'}
=a_{7.2} l_e^\mu + b_{7.2} p_p^\mu \,,
\end{equation}
whereas 
\begin{equation}
J_{8.3} = \int d\rho_{\gamma p} \frac{1}{(l_e\cdot k')(p_p\cdot k')} 
\end{equation}
and 
\begin{equation}
K_{8.3}^\mu = \int d\rho_{\gamma p} \frac{k'^\mu}{(l_e\cdot k')(p_p\cdot k')} 
=a_{8.3} k^\mu + b_{8.3}l_e^\mu + c_{8.3} p_p^\mu \,,
\end{equation}
so that 
\begin{eqnarray}
\nonumber && 
J_{8.3} \sim \frac{1}{M\omega} J_{7.2} \sim {\cal O}\left(\frac{1}{M^2}\right)  \quad ; \quad
a_{8.3} \sim 0 + {\cal O}\left(\frac{1}{M^3}\right)  \,, \\
&& 
b_{8.3} \sim  \frac{1}{M\omega} a_{7.2} + {\cal O}\left(\frac{1}{M^3}\right)  \quad ; \quad
c_{8.3} \sim \frac{1}{M\omega} b_{7.2} + {\cal O}\left(\frac{1}{M^4}\right)  
\label{asym7283}
\end{eqnarray}
as $M_p\to \infty$. 

The integrals in the remaining diagrams of 
Fig.~\ref{fig:allcuts} arise from cutting the electron
and proton lines to generate physical $ep\to ep\gamma$ or $ep\to ep$ scattering. 
The intermediate phase-space integrals in these cases are more complicated 
than those associated with the Compton cuts; fortunately, closed-form expressions 
for the integrals in the large $M_p$ limit 
suffice to leading order in the recoil expansion. 
With $P_0 \equiv p_p + l_e + k$, we note for diagram (5.2)
\begin{eqnarray}
\nonumber I_{5.2}&=&
\int \frac{d^3{\mathbf l}_e'}{2E_e'}\frac{d^3{\mathbf p}_p'}{2E_p'}
\delta^{(4)}(l_e'+p_p'-P_0) 
\equiv  \int d\rho_{ep\gamma} \\ 
&=&  \frac{\pi}{2 P_0^2} \sqrt{(P_0^2 - M_p^2 + m_e^2)^2 - 4 P_0^2 m_e^2} 
\sim \frac{\pi}{M} \sqrt{(E_e + \omega)^2 - m_e^2} \,
\end{eqnarray}
as ${M_p \to \infty}$. 
Moreover, 
\begin{equation}
J_{5.2}=\int d\rho_{ep\gamma} \frac{1}{(p_p' - p_p)^2} \,
\end{equation}
%Working in the large 
and 
\begin{equation}
J_{5.2} \sim \frac{\pi}{4M|\mathbf{l}_e + \mathbf{k}|} 
\log\left(
\frac{m_e^2 + l_e\cdot k -(E_e + \omega)^2 + \sqrt{(E_e+\omega)^2-m_e^2}\,|\mathbf{l}_e+\mathbf{k}|}
{m_e^2 + l_e\cdot k -(E_e + \omega)^2 - \sqrt{(E_e+\omega)^2-m_e^2}\,|\mathbf{l}_e+\mathbf{k}|}
\right) \,
\end{equation}
as $M_p\to \infty$. In addition, 
\begin{eqnarray}
K_{5.2}^\mu = \int d\rho_{ep\gamma} \frac{l_e'^\mu}{(p_p'-p_p)^2} = a_{5.2}P_{0e}^\mu +c_{5.2}p_p^\mu \,,
\end{eqnarray}
where $a_{5.2}$ and $c_{5.2}$ are given by the solution to 
\begin{eqnarray}
\nonumber (m_e^2+l_e\cdot k) J_{5.2} - \frac{I_{5.2}}{2} &=& a_{5.2} P_{0e}^2 + c_{5.2} p_p\cdot P_{0e} \,,\\
\nonumber p_p\cdot P_{0e} J_{5.2} + \frac{1}{2} I_{5.2} &=& a_{5.2} p_p\cdot P_{0e} + c_{5.2} M_p^2 \,,
\end{eqnarray}
so that in the large $M_p$ limit $a_{5.2} \sim 1/M$ and $c_{5.2} \sim 1/M^2$. 
Turning to the integrals from diagram $(6.1)$ we have 
\begin{equation}
I_{6.1}=\int d\rho_{ep\gamma}\frac{1}{(l_e'\cdot k)} \,,
\end{equation}
so that as ${M_p \to \infty}$ 
\begin{equation}
I_{6.1} \sim \frac{\pi}{2M\omega} 
\log\left(
\frac{E_e + \omega + \sqrt{(E_e+\omega)^2-m_e^2}}{E_e + \omega - \sqrt{(E_e+\omega)^2-m_e^2}}
\right) \,,
\end{equation}
as well as 
\begin{equation}
I_{6.1}'=\int d\rho_{ep\gamma}\frac{(p_p'-p_p)^2}{(l_e'\cdot k)} \,,
\end{equation}
where as ${M_p\to \infty}$
\begin{equation}
I_{6.1}' \sim 2(m_e^2 + l_e\cdot k) I_{6.1} - 2 I_{5.2} - 2\tilde I_{6.1}
\end{equation} 
with 
\begin{equation}
\!\!\!\!\!\!\!\tilde I_{6.1} = 
\frac{\pi \mathbf{k}\cdot \mathbf{l}_e }{M\omega^2} \left( 
\sqrt{(E_e+\omega)^2 - m_e^2} + 
\frac{(E_e + \omega)k\cdot l_e}{2  \mathbf{k}\cdot \mathbf{l}_e}
 \log \left( 
\frac{E_e + \omega + \sqrt{(E_e + \omega)^2 - m_e^2}}{E_e + \omega - \sqrt{(E_e + \omega)^2 - m_e^2}}
\right)
\right)\,.
\end{equation}
Moreover, 
\begin{equation}
J_{6.1}=\int d\rho_{ep\gamma} \frac{1}{(l_e'\cdot k)(p_p-p_p')^2} \,,
\end{equation}
so that as ${M_p\to \infty}$
\begin{equation}
J_{6.1} \sim \frac{\pi}{4M|\mathbf{l}_e|k\cdot l_e}\left(\log\left(\frac{A_+}{A_-}\right)-\log\left(\frac{B_{+}}{B_{-}}\right)\right) \,,
\end{equation}
where 
\begin{equation}
A_\pm=m_e^2+l_e\cdot k-(E_e+\omega)^2 
\pm|\mathbf{l}_e+\mathbf{k}|\sqrt{(E_e+\omega)^2-m_e^2}
\end{equation}
and
\begin{eqnarray}
\nonumber B_\pm &=& |\mathbf{l}_e|^2
(l_e\cdot k)^2-\left(\omega^2m_e^2-E_e\omega(l_e\cdot k)\right) A_{\pm}  \\
 && +|\mathbf{l}_e|(l_e\cdot k)\left((E_e+\omega)\omega|\mathbf{l}_e+\mathbf{k}|
\mp(\omega^2+\mathbf{l}_e\cdot\mathbf{k})\sqrt{(E_e+\omega)^2-m_e^2}\right) \,.
\end{eqnarray}
In addition, 
\begin{eqnarray}
K_{6.1}^\mu = \int d\rho_{ep\gamma} \frac{l_e'^\mu}{(l_e'\cdot k)(p_p-p_p')^2} 
=a_{6.1} l_e^\mu+b_{6.1}k^\mu+c_{6.1}p_p^\mu  \,,
\end{eqnarray}
where the undetermined coefficients are fixed by the solution to 
\begin{eqnarray}
\nonumber J_{5.2} &=& a_{6.1}l_e\cdot k + c_{6.1}p_p\cdot k \,,\\
\nonumber (m_e^2 + l_e\cdot k) J_{6.1}- \frac{I_{6.1}}{2} &=& a_{6.1}(m_e^2 + l_e\cdot k) + b_{6.1}l_e\cdot k + c_{6.1}p_p\cdot P_{0e}  \,, \\
\nonumber p_p\cdot P_{0e} J_{6.1}&=& a_{6.1}p_p\cdot l_e 
+ b_{6.1}p_p\cdot k + c_{6.1}M_p^2 \,, 
\end{eqnarray}
so that in the large $M_p$ limit $a_{6.1}, b_{6.1} \sim 1/M$ and $c_{6.1} \sim 1/M^2$. 
Also 
\begin{eqnarray}
\nonumber && L_{6.1}^{\mu\nu} = 
\int d\rho_{ep\gamma} \frac{l_e'^\mu l_e'^\nu}{(l_e'\cdot k)(p_p-p_p')^2} 
=d_{6.1}g^{\mu\nu}+e_{6.1}p_p^\mu p_p^\nu+f_{6.1}l_e^\mu l_e^\nu +g_{6.1} k^\mu k^\nu\\
\nonumber && +h_{6.1}(p_p^\mu l_e^\nu+l_e^\mu p_p^\nu)
+i_{6.1}(p_p^\mu k^\nu+ k^\mu p_p^\nu)+k_{6.1}(l_e^\mu k^\nu+ k^\mu l_e^\nu) \,,
\end{eqnarray}
where the undetermined coefficients are fixed by the solution to
\begin{eqnarray}
\nonumber && 4d_{6.1}+e_{6.1}M_p^2+f_{6.1}m_e^2+2h_{6.1}p_p\cdot l_e+2i_{6.1}p_p\cdot k
+2k_{6.1} l_e\cdot k=m_e^2 J_{6.1} \,,\\
\nonumber && d_{6.1} + e_{6.1} M_p^2 + h_{6.1}p_p\cdot l_e+ i_{6.1}p_p\cdot k=p_p\cdot P_{0e} c_{6.1} \,,\\
\nonumber && g_{6.1}p_p\cdot k +i_{6.1} M_p^2+k_{6.1} p_p\cdot l_e=p_p\cdot P_{0e} b_{6.1} \,,\\
\nonumber && f_{6.1} p_p\cdot l_e +h_{6.1} M_p^2+k_{6.1} p_p\cdot k=p_p\cdot P_{0e} a_{6.1} \,,\\
\nonumber && e_{6.1}p_p \cdot k+h_{6.1}l_e\cdot k=c_{5.2} \,,\\
\nonumber &&f_{6.1} l_e\cdot k +h_{6.1}p_p\cdot k=a_{5.2} \,,\\
\nonumber && d_{6.1}P_{0e}^2+e_{6.1}(p_p\cdot P_{0e})^2+f_{6.1}(l_e\cdot P_{0e})^2+g_{6.1}(l_e\cdot k)^2 + 2 h_{6.1}p_p\cdot P_{0e}l_e\cdot P_{0e} \\
\nonumber && +2 i_{6.1}p_p\cdot P_{0e}l_e\cdot k + 2 k_{6.1}l_e\cdot P_{0e}l_e\cdot k=(m_e^2+l_e\cdot k)^2 J_{6.1}-(m_e^2+l_e\cdot k) I_{6.1} + \frac{I_{6.1}'}{4} \,.
\end{eqnarray}

For the remaining $e-p-\gamma$ cuts we have
\begin{equation}
J_{7.1}=\int d\rho_{ep\gamma} \frac{1}{(l_e'-l_e)^2} \sim 
{\cal O}\left(\frac{1}{M}\right) 
\end{equation}
and 
\begin{equation}
K_{7.1}^\mu =\int d\rho_{ep\gamma} \frac{l_e'^\mu}{(l_e'-l_e)^2}  
=a_{7.1}l_e^\mu+b_{7.1}p_p^\mu  \,,
\end{equation}
whereas 
\begin{equation}
J_{8.1}=\int d\rho_{ep} 
\frac{1}{(p_p'\cdot k)(l_e'-l_e)^2} 
\end{equation}
and
\begin{equation}
K_{8.1}^\mu=\int d\rho_{ep} 
\frac{l_e'^\mu}{(p_p'\cdot k)(l_e'-l_e)^2}
=a_{8.1}l_e^\mu+b_{8.1}k^\mu+c_{8.1}p_p^\mu \,,
\end{equation}
so that 
\begin{eqnarray}
\nonumber && 
J_{8.1} \sim \frac{1}{M\omega} J_{7.1} \sim {\cal O}\left(\frac{1}{M^2}\right)  
\quad ; \quad
b_{8.1} \sim 0 + {\cal O}\left(\frac{1}{M^3}\right)  \,,
\\
&& 
a_{8.1} \sim  \frac{1}{M\omega} a_{7.1} + {\cal O}\left(\frac{1}{M^3}\right)  \quad ; \quad
c_{8.1} \sim \frac{1}{M\omega} b_{7.1} + {\cal O}\left(\frac{1}{M^4}\right)  \, 
\label{asym7181}
\end{eqnarray}
as $M_p\to \infty$. 

The integrals for the $e-p$ cuts follow from those we have just analyzed under 
the replacement of $P_0$ with $\tilde P_{0} \equiv l_e + p_p$. In this case, however, there
is an added complication because the integrals become infrared divergent when $p_p'=p_p$. 
This divergence cancels once we construct an observable quantity; nevertheless, we regulate
the integrals as they stand by adding a fictitious photon mass $m_\gamma^2$ -- this will allow
us to track the infrared divergences through the course of the calculation, 
so that we can demonstrate
the divergence cancellation manifestly. 
In what follows we set $m_\gamma^2$ to zero in all terms which are
finite in the $m_\gamma^2 \to 0$ limit. 
We have 
\begin{eqnarray}
\nonumber I_{8.2}&=&
\int \frac{d^3{\mathbf l}_e'}{2E_e'}\frac{d^3{\mathbf p}_p'}{2E_p'}\delta^{(4)}(l_e'+p_p'-\tilde P_0) 
\equiv  \int d\rho_{ep} \\
&& \sim \frac{\pi |\mathbf{l}_e|}{M}
\end{eqnarray} 
as $M_p \to \infty$. In addition, 
\begin{eqnarray}
J_{8.2}&=&\int d\rho_{ep} \frac{1}{p_p'\cdot k}\frac{1}{(p_p' - p_p)^2-m_\gamma^2} \\
\nonumber 
&\sim&\frac{\pi}{4 |\mathbf{l}_e | \omega M^2} \log \left( \frac{m_\gamma^2}{4 |\mathbf{l}_e|^2} \right)
\end{eqnarray}
as $M_p \to \infty$. 
Thus we see that $J_{8.2}$ vanishes in this limit save for the infrared divergent
piece, which we define as $J_{8.2}^{\rm div}$. 
In addition, 
\begin{eqnarray}
K_{8.2}^\mu = \int d\rho_{ep} \frac{1}{p_p'\cdot k}\frac{l_e'^\mu}{(p_p'-p_p)^2-m_\gamma^2} 
= a_{8.2}l_e^\mu + b_{8.2} k^\mu + c_{8.2}p_p^\mu \,. 
\end{eqnarray}
The coefficients are given by the solution to 
%$a_{8.2}$ and $c_{8.2}$ 
\begin{eqnarray}
\nonumber m_e^2 J_{8.2} - \frac{1}{2}\tilde I_{8.2}
&=& a_{8.2} m_e^2 + b_{8.2} l_e\cdot k + c_{8.2} p_p\cdot l_e \,,\\
\nonumber p_p\cdot l_e J_{8.2} +  \frac{1}{2}\tilde I_{8.2}
 &=& a_{8.2} p_p\cdot l_e + b_{8.2} p_p \cdot k + c_{8.2} M_p^2 \,, \\ 
\nonumber (l_e + p_p)\cdot k J_{8.2} -  I'_{8.2}
 &=& a_{8.2} l_e \cdot k +  c_{8.2} p_p\cdot k \,, 
\end{eqnarray}
where
\begin{equation}
\tilde I_{8.2}=\int d\rho_{ep} \frac{1}{p_p'\cdot k} \quad ; \quad
I'_{8.2} = \int d\rho_{ep} \frac{1}{(p_p' - p_p)^2-m_\gamma^2} \,.
\end{equation}
In the large $M_p$ limit we note that 
\begin{equation}
K_{8.2}^\mu \sim \frac{1}{M\omega} I'_{8.2} 
\end{equation}
so that $b_{8.2} \sim 0$, 
and we need only solve 
\begin{eqnarray}
\nonumber m_e^2 J_{8.2} - \frac{I_{8.2}}{2M\omega} 
&=& a_{8.2} m_e^2 + c_{8.2} M E_e \,,\\
  E_e J_{8.2} 
 &=& a_{8.2}  E_e + c_{8.2} M 
\end{eqnarray}
to determine the leading-order expressions for $a_{8.2}$ and $c_{8.2}$. 
We can track the infrared divergence in $J_{8.2}$ in $a_{8.2}$ and $c_{8.2}$ by solving
these equations with $I_{8.2}=0$ and $J_{8.2} = J_{8.2}^{\rm div}$, 
which yields $a_{8.2}^{\rm div}\sim J_{8.2}^{\rm div}$  and $c_{8.2}^{\rm div}\sim 0$
in leading order.

The integrals from diagram $(6.3)$ are 
\begin{eqnarray}
I_{6.3}&=&\int d\rho_{ep}\frac{1}{(l_e'\cdot k)} \\
\nonumber &\sim&\frac{\pi}{2 \omega M} \log \left( 
\frac{E_e + |\mathbf{l}_e |}{E_e - |\mathbf{l}_e |}\right)
\end{eqnarray}
as $M_p\to\infty$ and 
\begin{eqnarray}
I_{6.3}'&=&\int d\rho_{ep}\frac{(p_p'-p_p)^2}{(l_e'\cdot k)} \\
\nonumber &\sim& 2m_e^2 I_{6.3} - 2 \tilde I_{6.3} 
\end{eqnarray}
with 
\begin{equation}
\tilde I_{6.3} \sim \frac{\pi}{2M\omega} \left(
(E_e^2 - E_e |\mathbf{l}_e| \cos\theta_e) \log \left(\frac{E_e+|\mathbf{l}_e|}{E_e-|\mathbf{l}_e|}\right) 
+2 |\mathbf{l}_e|^2 \cos\theta_e
\right)
\end{equation}
as $M_p\to\infty$. 
We define 
$\mathbf{k} \cdot \mathbf{l}_e \equiv |\mathbf{k}| |\mathbf{l}_e| \cos\theta_{\rm e}$. 
Moreover,
\begin{eqnarray}
&& J_{6.3}=\int d\rho_{ep} 
\frac{1}{(l_e'\cdot k)}\frac{1}{(p_p-p_p')^2-m_{\gamma}^2} \\
\nonumber &\sim& 
\frac{\pi}{4 |\mathbf{l}_e | (l_e\cdot k) M} \left( \log \frac{m_\gamma^2}{4 |\mathbf{l}_e|^2} +
\log \frac{m_e^2\omega^2}{ (l_e\cdot k)^2}
\right)
\end{eqnarray}
as $M_p\to\infty$. In this case we see that $J_{6.3}$ has both infrared finite and divergent pieces in the
$M_p\to\infty$ limit -- the latter we define as $J_{6.3}^{\rm div}$. 
%Thus we see that $J_{8.2}$ vanishes in this limit save for the infrared divergent
%piece, which we define as 
%
Finally
\begin{eqnarray}
K_{6.3}^\mu = \int d\rho_{ep} \frac{1}{(l_e'\cdot k)}\frac{l_e'^\mu}{(p_p-p_p')^2-m_{\gamma}^2} 
=a_{6.3} l_e^\mu+b_{6.3}k^\mu+c_{6.3}p_p^\mu  \,,
\end{eqnarray}
where the undetermined coefficients are fixed by the solution to 
\begin{eqnarray}
\nonumber p_p\cdot k J_{8.2}&=& a_{6.3}l_e\cdot k + c_{6.3}p_p\cdot k \,,\\
\nonumber m_e^2 J_{6.3}- \frac{I_{6.3}}{2} &=& a_{6.3}m_e^2 + b_{6.3}l_e\cdot k + c_{6.3}p_p\cdot l_e  \,, \\
\nonumber p_p\cdot l_e J_{6.3}&=& a_{6.3}p_p\cdot l_e + b_{6.3}p_p\cdot k + c_{6.3}M_p^2 \,. 
\end{eqnarray}
Also 
\begin{eqnarray}
\nonumber && L_{6.3}^{\mu\nu} = 
\int d\rho_{ep} \frac{1}{(l_e'\cdot k)}\frac{l_e'^\mu l_e'^\nu}{(p_p-p_p')^2-m_{\gamma}^2} 
=d_{6.3}g^{\mu\nu}+e_{6.3}p_p^\mu p_p^\nu+f_{6.3}l_e^\mu l_e^\nu +g_{6.3} k^\mu k^\nu\\
\nonumber && +h_{6.3}(p_p^\mu l_e^\nu+l_e^\mu p_p^\nu)
+i_{6.3}(p_p^\mu k^\nu+ k^\mu p_p^\nu)+k_{6.3}(l_e^\mu k^\nu+ k^\mu l_e^\nu) \,,
\end{eqnarray}
where the undetermined coefficients are fixed by the solution to
\begin{eqnarray}
\nonumber && 4d_{6.3}+e_{6.3}M_p^2+f_{6.3}m_e^2+2h_{6.3}p_p\cdot l_e+2i_{6.3}p_p\cdot k
+2k_{6.3} l_e\cdot k=m_e^2 J_{6.3} \,,\\
\nonumber && d_{6.3} + e_{6.3} M_p^2 + h_{6.3}p_p\cdot l_e+ i_{6.3}p_p\cdot k=p_p\cdot l_e c_{6.3} \,,\\
\nonumber && g_{6.3}p_p\cdot k +i_{6.3} M_p^2+k_{6.3} p_p\cdot l_e=p_p\cdot l_e b_{6.3} \,,\\
\nonumber && f_{6.3} p_p\cdot l_e +h_{6.3} M_p^2+k_{6.3} p_p\cdot k=p_p\cdot l_e a_{6.3} \,,\\
\nonumber && e_{6.3}p_p \cdot k+h_{6.3}l_e\cdot k=p_p\cdot k c_{8.2} \,,\\
\nonumber &&f_{6.3} l_e\cdot k +h_{6.3}p_p\cdot k=p_p\cdot k a_{8.2} \,,\\
\nonumber && d_{6.3}m_e^2+e_{6.3}(p_p\cdot l_e)^2+f_{6.3}m_e^4+g_{6.3}(l_e\cdot k)^2 + 2 h_{6.3}p_p\cdot l_e m_e^2 +2 i_{6.3}p_p\cdot l_e l_e\cdot k\\
\nonumber && + 2 k_{6.3}m_e^2 l_e\cdot k=m_e^4 J_{6.3}-m_e^2 I_{6.3} + \frac{I_{6.3}'}{4} \,.
\end{eqnarray}

We can track the infrared divergence in $J_{6.3}$ in the solutions for the vector and
tensor coefficients by solving
the equations in the large $M_p$ limit with $I_{6.3} \sim I_{6.3}'\sim 0$ 
and $J_{6.3} \sim J_{6.3}^{\rm div}$, with $a_{8.2} \sim  a_{8.2}^{\rm div}$, 
which yields $a_{6.3}^{\rm div} \sim f_{6.3}^{\rm div} \sim J_{6.3}^{\rm div}$ 
with all other coefficients zero in this limit.

\section{${\overline{|{\cal M}|^2}_{\rm \ot \ odd}}$ in Leading Order}

In what follows we report the contributions 
to the \ot-odd correlation in ${\cal O}(\alpha)$ up to corrections of recoil order. 
We organize the results as per 
the various gauge-invariant families 
we describe in the main body of the text, 
employing the subscript convention which 
follows the labeling in Figs.~\ref{fig:tree} and \ref{fig:allcuts}. 
We use the integrals and Passarino-Veltman coefficients defined in Appendix A. 
The result for the $\gamma-e$ family is 
\begin{eqnarray}
\nonumber && 
{\overline{|{\cal M}|^2}_{\rm \ot \ odd}} \left[1.01+1.02+2.01+2.02+
5.1.01+5.1.02+6.2.01+6.2.02\right] \\
\nonumber && = -\alpha^2 g_V^2 G_F^2 \xi 64 M^2 (1-\lambda^2)\Bigg(
\frac{m_e^2}{(l_e\cdot k)^2 \omega} a_1 + 
\frac{m_e^2}{(l_e\cdot k)^2 \omega} J_1 + 
\frac{1}{l_e\cdot k\,\omega} c_2 + \frac{1}{l_e\cdot k\,\omega} a_1 
- \frac{1}{l_e\cdot k \,\omega} J_1 \\
\nonumber && + \frac{m_e^2}{l_e\cdot k\,\omega} b_2+ \frac{m_e^2}{l_e\cdot k\,\omega} a_2 
 - \frac{m_e^2}{l_e\cdot k\,\omega} J_2 +  \frac{ M E_e}{\omega} k_{6.2} 
+  \frac{M E_e}{\omega} g_{6.2} -  \frac{M E_e}{\omega} b_{6.2} 
-  \frac{2 M E_e}{\omega} a_{6.2}    
\\
\nonumber && 
+  \frac{M E_e}{\omega} J_{6.2}  
+  \frac{M E_e}{l_e\cdot k\,\omega}  b_{5.1} 
-\frac{M E_e}{l_e\cdot k \,\omega}  J_{5.1} 
-\frac{M E_e}{2 l_e\cdot k}  g_{6.2} 
+\frac{M E_e}{2 l_e\cdot k}  f_{6.2} 
+\frac{M E_e}{l_e\cdot k}  a_{6.2} 
+ \frac{M^2}{\omega} i_{6.2}  
\\
\nonumber && 
- \frac{M^2 }{2\omega} c_{6.2} - \frac{M^2 E_e}{l_e\cdot k} c_{6.2} 
+\frac{ M^2 E_e}{l_e\cdot k} h_{6.2} 
 + \frac{M^2 }{2 l_e\cdot k\,\omega}  a_{5.1}  
 +  \frac{M^3}{2 l_e\cdot k} e_{6.2}\Bigg) \,.
\end{eqnarray}
The result for the $\gamma-p$ family is 
\begin{eqnarray}
\nonumber && 
{\overline{|{\cal M}|^2}_{\rm \ot \ odd}} \left[3.01+3.02+4.01+4.02+
7.2.01+7.2.02+8.3.01+8.3.02\right] \\
\nonumber && = -\alpha^2 g_V^2 G_F^2 \xi 64 M^3 (1-\lambda^2)\Bigg(
\frac{E_e}{l_e\cdot k\, \omega} a_{7.2}  
+ \frac{E_e}{l_e\cdot k\, \omega} J_{7.2} 
- \frac{1}{l_e\cdot k\, \omega^2} J_{3}  
- \frac{M E_e}{l_e\cdot k} b_{8.3} 
- \frac{M}{\omega} a_{8.3}  \\
\nonumber && 
+ \frac{ME_e}{l_e\cdot k} a_{8.3} 
- \frac{ME_e}{l_e\cdot k} J_{8.3}  
+ \frac{M}{2l_e\cdot k\, \omega} b_{7.2} 
+ \frac{M}{l_e\cdot k\, \omega} J_{4} 
- \frac{M^2}{2 l_e\cdot k} c_{8.3} \Bigg) \\
\nonumber && 
= 0 + {\cal O}(M)\,, 
\end{eqnarray}
where we employ Eqs.~(\ref{asym34}) and (\ref{asym7283}) to determine
that the contribution to this family vanishes in leading order in $M$. 
The results for the $e-p-\gamma$ families are 
\begin{eqnarray}
\nonumber && 
{\overline{|{\cal M}|^2}_{\rm \ot \ odd}} \left[5.2.01+5.2.02+6.1.01+6.1.02\right] \\
\nonumber && = -\alpha^2 g_V^2 G_F^2 \xi 64 M^3 (1-\lambda^2)\Bigg(
\frac{2 E_e}{\omega} k _{6.1}+ \frac{2 M }{\omega} i _{6.1} - \frac{ M }{\omega} c _{6.1} - \frac{2 E_e}{l_e\cdot k} \frac{1}{\omega} a _{5.2} -\frac{2 m_e^2}{l_e\cdot k}  k _{6.1} \\
\nonumber && + \frac{ m_e^2}{l_e\cdot k} f _{6.1} + \frac{m_e^2 }{l_e\cdot k} J _{6.1} - \frac{ M }{l_e\cdot k}\frac{1}{\omega} c _{5.2}  -  \frac{2 M E_e}{l_e\cdot k} i _{6.1} +  \frac{2M E_e }{l_e\cdot k} h _{6.1} + \frac{2M E_e  }{l_e\cdot k} c _{6.1} +  \frac{ M^2}{l_e\cdot k} e _{6.1}\Bigg) \,
\end{eqnarray}
and 
\begin{eqnarray}
\nonumber && 
{\overline{|{\cal M}|^2}_{\rm \ot \ odd}} \left[7.1.01+7.1.02+8.1.01+8.1.02\right] \\
\nonumber && = -\alpha^2 g_V^2 G_F^2 \xi 64 M^3 (1-\lambda^2)\Bigg(
\frac{2 E_e}{l_e\cdot k\,\omega} a _{7.1} + \frac{M}{l_e\cdot k\,\omega} b _{7.1}
- \frac{2 M E_e}{l_e\cdot k} a _{7.1} - \frac{2M}{\omega} b _{7.1} \quad\quad\quad \quad
\\
\nonumber && 
+ \frac{2ME_e}{l_e\cdot k} b _{7.1} 
-  \frac{M^2}{l_e\cdot k} c _{8.1} \Bigg) \\
\nonumber && 
= 0 + {\cal O}(M)\,, 
\end{eqnarray}
where we employ Eq.~(\ref{asym7181}) to determine
that the contribution to this family vanishes in leading order in $M$. 
We emphasize that the contributions which vanish do so simply 
to the order of the recoil expansion in which we work. 
Finally, the result for the $e-p$ family is
\begin{eqnarray}
\nonumber && 
{\overline{|{\cal M}|^2}_{\rm \ot \ odd}} \left[6.3.01+6.3.02+8.2.01+8.2.02\right] \\
\nonumber && = -\alpha^2 g_V^2 G_F^2 \xi 64 M^3 (1-\lambda^2)
\Bigg( \frac{2 m_e^2 }{l_e\cdot k}k _{6.3}-\frac{2 E_e }{\omega} k _{6.3}
-\frac{2 E_e }{\omega} a _{6.3} -\frac{2 M }{\omega} i _{6.3} 
-\frac{ M }{\omega} c _{6.3} - \frac{ m_e^2}{l_e\cdot k} f _{6.3} \\  
\nonumber && +\frac{ 2m_e^2 }{l_e\cdot k} a _{6.3} -  \frac{ m_e^2}{l_e\cdot k} J _{6.3} 
+\frac{2 M E_e }{l_e\cdot k} a _{8.2} + \frac{2M E_e }{l_e\cdot k} i _{6.3} 
-\frac{2  M E_e }{l_e\cdot k}h _{6.3}   + \frac{ M^2}{l_e\cdot k} c _{8.2} 
-  \frac{ M^2 }{l_e\cdot k}e _{6.3}
\Bigg)\,. 
\end{eqnarray} 
From Appendix A we note that 
$a_{8.2}^{\rm div}\sim J_{8.2}^{\rm div} \sim (l_e\cdot k)J_{6.3}^{\rm div}/(M\omega)$  and 
$a_{6.3}^{\rm div} \sim f_{6.3}^{\rm div} \sim J_{6.3}^{\rm div}$ 
with all other coefficients zero in leading order in the recoil expansion. 
Thus we see explicitly that the infrared divergence really does cancel in 
${\cal O}(M^2)$.
%leading order. 


\begin{thebibliography}{99}

\bibitem{Jackson:1957zz} 
  J.~D.~Jackson, S.~B.~Treiman, and H.~W.~Wyld,
  %``Possible tests of time reversal invariance in Beta decay,''
  Phys.\ Rev.\  {\bf 106}, 517 (1957).
  %%CITATION = PHRVA,106,517;%%

\bibitem{sachs} R.~G. Sachs, {\em The Physics of Time Reversal}
(University of Chicago Press, Chicago, 1985), p. 112ff.

%\cite{Harvey:2007rd}
\bibitem{Harvey:2007rd} 
  J.~A.~Harvey, C.~T.~Hill, and R.~J.~Hill,
  %``Anomaly mediated neutrino-photon interactions at finite baryon density,''
  Phys.\ Rev.\ Lett.\  {\bf 99}, 261601 (2007). 
%  [arXiv:0708.1281 [hep-ph]].
  %%CITATION = ARXIV:0708.1281;%%

%\cite{Harvey:2007ca}
\bibitem{Harvey:2007ca} 
  J.~A.~Harvey, C.~T.~Hill, and R.~J.~Hill,
  %``Standard Model Gauging of the Wess-Zumino-Witten Term: Anomalies, Global Currents and pseudo-Chern-Simons Interactions,''
  Phys.\ Rev.\ D {\bf 77}, 085017 (2008).
%  [arXiv:0712.1230 [hep-th]].
  %%CITATION = ARXIV:0712.1230;%%

%\cite{Hill:2009ek}
\bibitem{Hill:2009ek} 
  R.~J.~Hill,
  %``Low energy analysis of nu N ---> nu N gamma in the Standard Model,''
  Phys.\ Rev.\ D {\bf 81}, 013008 (2010). 
%  [arXiv:0905.0291 [hep-ph]].
  %%CITATION = ARXIV:0905.0291;%%

\bibitem{svgdh1} S. Gardner and D. He, in preparation.

\bibitem{Mumm:2011nd} 
  H.~P.~Mumm, T.~E.~Chupp, R.~L.~Cooper, K.~P.~Coulter, S.~J.~Freedman, B.~K.~Fujikawa, A.~Garcia, and G.~L.~Jones {\it et al.},
  %``A New Limit on Time-Reversal Violation in Beta Decay,''
  Phys.\ Rev.\ Lett.\  {\bf 107}, 102301 (2011). 
%  [arXiv:1104.2778 [nucl-ex]].
  %%CITATION = ARXIV:1104.2778;%%

\bibitem{calaprice}
  A.~L.~Hallin, F.~P.~Calaprice, D.~W.~MacArthur, L.~E.~Piilonen, M.~B.~Schneider, and D.~F.~Schreiber,
  %``Test Of Time Reversal Symmetry In The Beta Decay Of Ne-19,''
  Phys.\ Rev.\ Lett.\  {\bf 52}, 337 (1984). 
  %%CITATION = PRLTA,52,337;%%

\bibitem{Callan:1967zz} 
  C.~G.~Callan and S.~B.~Treiman,
  %``Electromagnetic Simulation of T Violation in Beta Decay,''
  Phys.\ Rev.\  {\bf 162}, 1494 (1967).
  %%CITATION = PHRVA,162,1494;%%

\bibitem{Ando:2009jk} 
  S.~-i.~Ando, J.~A.~McGovern, and T.~Sato,
  %``The D coefficient in neutron beta decay in effective field theory,''
  Phys.\ Lett.\ B {\bf 677}, 109 (2009). 
%  [arXiv:0902.1194 [nucl-th]].
  %%CITATION = ARXIV:0902.1194;%%

%\cite{Braguta:2001nz}
\bibitem{braguta}
  V.~V.~Braguta, A.~A.~Likhoded, and A.~E.~Chalov,
  %``T odd correlation in the K(l3 gamma) decay,''
  Phys.\ Rev.\ D {\bf 65}, 054038 (2002);
  Yad.\ Fiz.\  {\bf 65}, 1920 (2002) 
  [Phys.\ Atom.\ Nucl.\  {\bf 65}, 1868 (2002)].
%  [hep-ph/0106147].
  %%CITATION = HEP-PH/0106147;%%

%\cite{Khriplovich:2010rz}
\bibitem{khriprud}
  I.~B.~Khriplovich and A.~S.~Rudenko,
%``$K^+_{l3\gamma}$ decays revisited: branching ratios and T-odd momenta correlations,''
  Phys.\ Atom.\ Nucl.\  {\bf 74}, 1214 (2011). 
%  [arXiv:1012.0147 [hep-ph]].
  %%CITATION = ARXIV:1012.0147;%%

%\cite{Cutkosky:1960sp}
\bibitem{Cutkosky:1960sp} 
  R.~E.~Cutkosky,
  %``Singularities and discontinuities of Feynman amplitudes,''
  J.\ Math.\ Phys.\  (N.Y.)\ {\bf 1}, 429 (1960).
  %%CITATION = JMAPA,1,429;%%

%\cite{Okun:1967ww}
\bibitem{Okun:1967ww} 
  L.~B.~Okun and I.~B.~Khriplovich,
  %``C odd correlation in the K0(mu3) decay and the pion electromagnetic form-factor,''
  Yad.\ Fiz.\  {\bf 6}, 821 (1967)

  [Sov.\ J.\ Nucl.\ Phys.\  {\bf 6}, 598 (1968)].
  %%CITATION = SJNCA,6,598;%%

\bibitem{Low:1958sn} 
  F.~E.~Low,
  %``Bremsstrahlung of very low-energy quanta in elementary particle collisions,''
  Phys.\ Rev.\  {\bf 110}, 974 (1958).
  %%CITATION = PHRVA,110,974;%%

\bibitem{Muller:2006gu} 
  E.~H.~M\"uller, B.~Kubis, and U.~-G.~Mei{\ss}ner,
  %``T-odd correlations in radiative K(l3)+ decays and chiral perturbation theory,''
  Eur.\ Phys.\ J.\ C {\bf 48}, 427 (2006).
%  [hep-ph/0607151].
  %%CITATION = HEP-PH/0607151;%%

\bibitem{PS} 
M.~E.~Peskin and D.~V.~Schroeder, {\em An Introduction to Quantum Field Theory}
(Addison-Wesley, Reading, MA, 1995). 

\bibitem{gaponov} 
  Y. V.~Gaponov and R.~U.~Khafizov,
%``Study of radiative neutron beta decay,''
 Yad.\ Fiz.\  {\bf 59}, 1270 (1996)
 [Phys.\ Atom.\ Nucl.\  {\bf 59}, 1213 (1996)]; 
 Phys.\ Lett.\ B {\bf 379}, 7 (1996); 
 Nucl.\ Instrum.\ Methods Phys. Res., Sect. A {\bf 440}, 557 (2000). 
  %%CITATION = PANUE,59,1213;%%

\bibitem{bgmz} 
  V.~Bernard, S.~Gardner, U.-G.~Mei{\ss}ner, and C.~Zhang,
%``Radiative neutron beta decay in effective field theory,''
  Phys.\ Lett.\ B {\bf 593}, 105 (2004);
  599, 348(E) (2004). 
%  [Erratum-ibid.\ B {\bf 599}, 348 (2004)]. 
%  [hep-ph/0403241].
  %%CITATION = HEP-PH/0403241;%%
Here we choose a different phase convention for ${\cal M}_0$ so that
no $i$ appears. 

\bibitem{nico} 
J.~S.~Nico, M.~S.~Dewey, T.~R.~Gentile, H.~P.~Mumm, A.~K.~Thompson, B.~M.~Fisher, I.~Kremsky, and F.~E.~Wietfeldt {\it et al.},
%``Observation of the radiative decay mode of the free neutron,''
 Nature (London) {\bf 444}, 1059 (2006).
 %%CITATION = NATUA,444,1059;%%

\bibitem{cooper} 
  R.~L.~Cooper, T.~E.~Chupp, M.~S.~Dewey, T.~R.~Gentile, H.~P.~Mumm, J.~S.~Nico, A.~K.~Thompson, and B.~M.~Fisher {\it et al.},
  %``Radiative beta decay of the free neutron,''
  Phys.\ Rev.\ C {\bf 81}, 035503 (2010).
  %%CITATION = PHRVA,C81,035503;%%
Note Fig.~11 for a comparison with the theoretical photon energy spectrum. 

\bibitem{KLN} T.~Kinoshita, J.\ Math.\ Phys.\ (N.Y.) {\bf 3}, 650 (1962); 
  T.~D.~Lee and M.~Nauenberg, Phys.\ Rev.\ {\bf 133}, B1549 (1964).
%Kinoshita (1960) Lee Nauenberg (1964) 
%T.~Kinoshita, J.\ Math.\ Phys.\ {\bf 3}, 650 (1962); 
%T.~D.~Lee and M.~Nauenberg, Phys.\ Rev.\ {\bf 133}, 1549 (1964). 

\bibitem{passvelt} G.~Passarino 
and M.~J.~G.~Veltman, Nucl.\ Phys.\ {\bf B160}, 151 (1979). 

\bibitem{vermaseren} J.~A.~M.~Vermaseren, arXiv:math-ph/0010025. 
Note also http://www.nikhef.nl/$\sim$form/. 

\bibitem{byrne} J.~Byrne, R.~U.~Khafizov, Yu.~A.~Mostovoi, 
O.~Rozhunov, V.~A.~Solovei, M.~Beck, V.~U.~Kozlov, and N.~Severijns, 
J.\ Res.\ Natl.\ Inst.\ Stand.\ Techno.\ {\bf 110}, 415 (2005).

\bibitem{pdg} 
K.~Nakamura {\it et al.} (Particle Data Group), J.\ Phys.\ G {\bf 37}, 075021 (2010), 
and 2011 partial update for the 2012 edition. Note http://pdg.lbl.gov$\,.$

%\cite{Wildenthal:1983zz}
\bibitem{Wildenthal:1983zz} 
  B.~H.~Wildenthal, M.~S.~Curtin, and B.~A.~Brown,
  %``Predicted features of the beta decay of neutron-rich sd-shell nuclei,''
  Phys.\ Rev.\ C {\bf 28}, 1343 (1983).
  %%CITATION = PHRVA,C28,1343;%%

\bibitem{caurier}
E.~Caurier, A.~P.~Zuker, A.~Poves, and G.~Martinez-Pinedo, Phys.\ Rev.\ C {\bf 50}, 
225 (1994).

\bibitem{quench}
E.~Caurier, A.~Poves, and A.~P.~Zuker, Phys.\ Rev.\ Lett.\ {\bf 74}, 1517 (1995). 

\bibitem{shimizu} 
F.~Shimizu, K.~Shimizu, and H.~Takuma, Phys.\ Rev.\ A {\bf 39}, 2758 (1989). 

%\bibitem{nucdata}
%http://www.nndc.bnl.gov/$\,$.

\bibitem{babrown} 
B.~A.~Brown and B.~H.~Wildenthal, Phys.\ Rev.\ C {\bf 28}, 2397 (1983). 

\bibitem{brownWild}
B.~A.~Brown and B.~H.~Wildenthal, At.\ Data Nucl.\ Data Tables {\bf 33}, 347 (1985). 

\bibitem{holstein} 
B.~R.~Holstein, Rev.\ Mod.\ Phys.\ {\bf 46}, 789 (1974); 48, 673(E) (1976). 

\bibitem{svgdh2} S. Gardner and D. He, in preparation.

\end{thebibliography}
\end{document}